\documentclass[aps,prl,reprint,superscriptaddress]{revtex4-1}

\usepackage{soul}
\usepackage{graphicx}
\usepackage{amsmath,amssymb,amsfonts}
\usepackage{enumitem}
\usepackage{xcolor}
\usepackage{soul,xcolor}
\usepackage{tikz,tkz-euclide}
\def\DJo{$\;$\kern-.4em \hbox{D\kern-.8em\raise.15ex\hbox{--}\kern.35em okovi\'c}}
\def\CC{{\rm\kern.24em \vrule width.04em height1.46ex depth-.07ex
\kern-.30em C}}
\def\P{{\rm I\kern-.25em P}}
\def\NN{{\rm I\kern-.15em N}}
\def\RR{{\rm
         \vrule width.04em height1.57ex depth-.0ex
         \kern-.03em R}}
\def\RR{{\rm I\kern-.23em R}}
\def\id{{\rm 1\kern-.22em l}}
\def\ZZ{{\sf Z\kern-.44em Z}}

\newtheorem{psatz}{Satz}[section]
\newtheorem{pdef}{Definition}[section]
\newtheorem{conjecture}{Vermutung}[section]

\newenvironment{eqblock}[2]{\beq\label{#2}\begin{array}{#1}}{\end{array}
                                \eeq}
\newenvironment{neqblock}[1]{\[\begin{array}{#1}}{\end{array}\]}
\newcommand{\beqb}{\begin{eqblock}}
\newcommand{\eeqb}{\end{eqblock}} 
\newcommand{\nbeqb}{\begin{neqblock}}
\newcommand{\neeqb}{\end{neqblock}}

\newcommand{\beq}{\begin{equation}}
\newcommand{\beqa}{\begin{eqnarray}}
\newcommand{\eeq}{\end{equation}}
\newcommand{\eeqa}{\end{eqnarray}}
\newcommand{\nbeqa}{\begin{eqnarray*}}
\newcommand{\neeqa}{\end{eqnarray*}}

\newcommand{\ket}[1]{| #1 \rangle}

\usepackage[colorlinks=true,citecolor=blue,linkcolor=blue,urlcolor=blue]{hyperref}

\begin{document}
\title{Probe for bound states of SU(3) fermions and colour deconfinement}

\author{Wayne J. Chetcuti}
\affiliation{Dipartimento di Fisica e Astronomia ``Ettore Majorana'', Via S. Sofia 64, 95127 Catania, Italy}
\affiliation{INFN-Sezione di Catania, Via S. Sofia 64, 95127 Catania, Italy}
\affiliation{Quantum Research Center, Technology Innovation Institute, Abu Dhabi, P.O. Box 9639, UAE}
\thanks{wayne.chetcuti@dfa.unict.it}

\author{Juan Polo}
\affiliation{Quantum Research Center, Technology Innovation Institute, Abu Dhabi, P.O. Box 9639, UAE}

\author{Andreas Osterloh}
\affiliation{Quantum Research Center, Technology Innovation Institute, Abu Dhabi, P.O. Box 9639, UAE}

\author{Paolo Castorina}
\affiliation{INFN-Sezione di Catania, Via S. Sofia 64, 95127 Catania, Italy}
\affiliation{Institute of Particle and Nuclear Physics, Charles University, Prague, Czech Republic}

\author{Luigi Amico}
\thanks{On leave from the Dipartimento di Fisica e Astronomia ``Ettore Majorana'', University of Catania.}
\affiliation{INFN-Sezione di Catania, Via S. Sofia 64, 95127 Catania, Italy}
\affiliation{Quantum Research Center, Technology Innovation Institute, Abu Dhabi, P.O. Box 9639, UAE}
\affiliation{Centre for Quantum Technologies, National University of Singapore, 3 Science Drive 2, Singapore 117543, Singapore}
\affiliation{LANEF `Chaire d’excellence’, Université Grenoble-Alpes \& CNRS, F-38000 Grenoble, France}

\begin{abstract}
Fermionic artificial matter realised with cold atoms grants access to an unprecedented degree of control  on sophisticated many-body effects with an enhanced flexibility of the operating conditions. We consider three-component fermions with attractive interactions to study the formation of complex bound states, whose nature goes beyond the standard fermion pairing occurring in quantum materials. Such systems display clear analogies with quark matter. 
Here, we address the nature of the bound states of a three-component fermionic system in a ring-shaped trap through the persistent current. In this way, we demonstrate that we can distinguish between color superfluid and trionic bound states. By analyzing finite temperature effects, we show how finite temperature can lead to the deconfinement of bound states.
For weak interactions, the deconfinement occurs because of scattering states. In this regime, the deconfinement depends on the trade-off between interactions and thermal fluctuations.
For strong interactions the features of the persistent current result from the properties of a suitable gas of bound states.
\end{abstract}
\date{\today}

\maketitle

{\it Introduction --}\label{intro} Mutually attracting quantum many-body systems can form bound states. Their nature depends on the particles' quantum statistics. Bosons can give rise to `bright solitons' in which all the particles  are bound together~\cite{strecker2002formation,Kanamoto2005symmetry,calabrese2007dynamics,naldesi2019rise}.
Due to the Pauli exclusion principle, such states are hindered for fermions.
Nevertheless, two-component fermions with opposite spin can form bounded pairs~\cite{leggett2006quantum}.

The advent of ultracold atomic systems has enabled the investigation of many-body systems made of interacting $N$-component fermions in the laboratory: the so-called SU($N$) fermions~\cite{scazza2014observation,scazza2016direct,av2010two,cazalilla2014ultracold,capponi2016phases,sonderhouse2020thermodynamics}. In contrast to their two-component counterparts,  $N$-component fermions can form  bound states of different type and nature. 
Here, we focus on SU($3$) fermions. On one hand, this can provide paradigmatic features of the bound states that can be formed for the general cases of $N\!>\!2$. On the other, three-component fermions are of special interest because of their  potential to mimic  quarks and specific aspects of  quantum chromodynamics (QCD)~\cite{Greensite2011,Cherng2007superfluidity,Rapp2007color,Klingschat2010exact}
for which it is clearly advantageous to explore ``low-energy" quantum analogues~\cite{baym2010bcs,rico2018so,banerjee2012atomic,tajima2021cooper,dalmonte2016lattice}.
Specifically, SU($3$) fermions can form two types of bound states: a colour superfluid (CSF) wherein  two colours are paired,
and the other  is unpaired; and a  trion where all colours are involved in the bound state. Trions and CSFs are the analogues of hadrons and quark pairs in QCD. As such, important aspects of the QCD phase diagram like colour deconfinement and resonance formation in nuclear matter, can be analysed in cold atom platforms. 

Fueled by the recent aforementioned research activity in quantum technology, a considerable interest has been devoted to three-component fermions ~\cite{honerkamp2004bcs,Cherng2007superfluidity,Rapp2007color,catelani2008phase,Capponi2008molecular,Klingschat2010exact,Batchelor2010exact,kuhn2012phase,pohlmann2013trion,Guan2013Fermi}. However, devising physical observables paving the way to explore the nature of the SU($3$) bound states in cold atoms systems  remains a challenging problem.        

In this paper, we demonstrate how  the frequency of the persistent current in a ring-shaped gas of three-component fermions pierced by an effective magnetic field, can provide the sought-after observable to study the problem. 
The persistent current is a matter-wave current probing the phase coherence of the system~\cite{imry2002intro}, that in line with the recent research activity in atomtronics~\cite{amico2020roadmap,amico2021atomtronic}, can be exploited as a diagnostic tool to explore quantum states. Persistent currents have been observed experimentally for both bosonic~\cite{phillips2007observation,ramanathan2011superflow,pandey2019hyper,wolf2021stationary} and very recently fermionic systems~\cite{kevin2022persistent,roati2022quantum}.
\begin{figure}[h!]
\centering
    \includegraphics[width=1\linewidth]{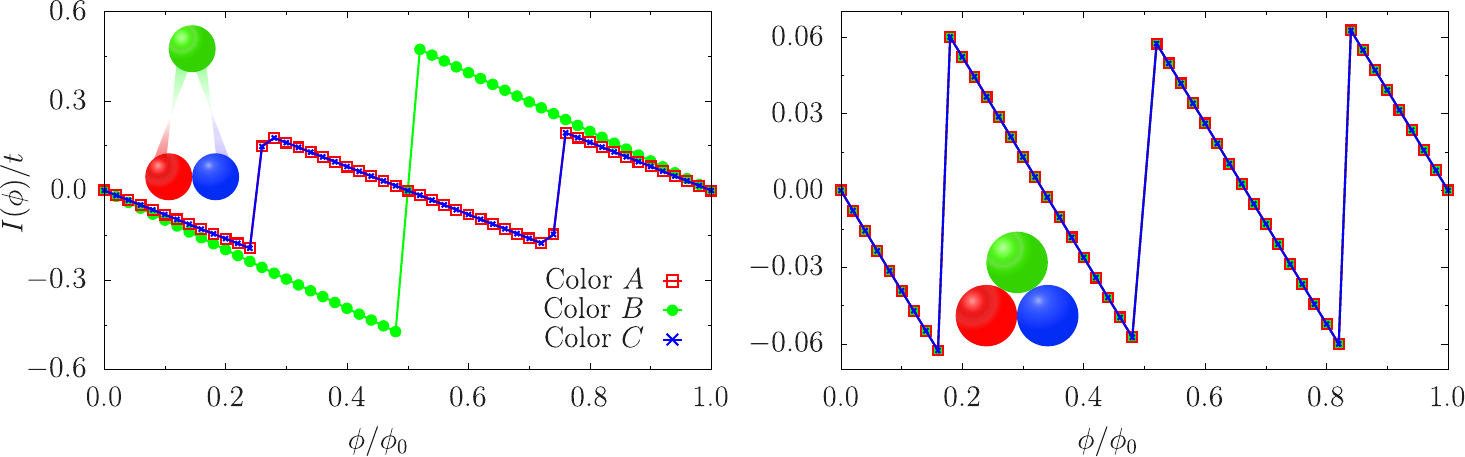}%
    \put(-139,67){(\textbf{a})}
    \put(-16,67){(\textbf{b})}
    \caption{Persistent current $I(\phi)$ of the three colours against the effective magnetic flux $\phi/\phi_{0}$. The left (right) figure depicts the persistent current of a CSF (trion). The interactions for the CSF are $|U_{AB}|/t=|U_{BC}|/t = 0.01$ and $|U_{AC}|/t=3$. For a trion, $|U|/t = 3$ for all colours. All presented results are obtained for $N_{p} = 9$, $L = 15$ and  using DMRG. The lines are meant to be a guide to the eye for the reader, to aid in perceiving the fractionalization.  }
    \label{fig:main}
\end{figure}

Very important for our approach is the Leggett theorem, stating that the persistent current periodicity is dictated by the system's {\it effective} flux quantum~\cite{leggett1991}. For example, the effective flux quantum of a gas of non-interacting particles is the bare flux $\phi_{0}$; while in a gas of Cooper pairs, the period is halved since a flux quantum is shared by two particles~\cite{byers1961theoretical,onsager1961magnetic}. A persistent current with a periodicity reduced by the total number of particles $1/N_{p}$ has been found, indicating the formation of an $N_{p}$-bound state in bosonic systems~\cite{polo2020exact,naldesi2020enhancing,polo2020quantum}. Recently, the persistent current was used to investigate an SU($N$) fermionic atomtronic circuit with repulsive interactions~\cite{chetcuti2020persistent,andrea2021interaction}.

In our work, trion and CSF bound states correspond to specific ways in which the persistent current responds to the effective magnetic field. By monitoring the persistent current for different interaction regimes, we  demonstrate how thermal fluctuations can lead to a specific deconfinement of the bound states. As an experimental probe in the cold atoms quantum technology, we analyse the time-of-flight imaging~\cite{amico2020roadmap,amico2021atomtronic,ramanathan2011superflow,beattie2013persistent,lewenstein2012ultracold}.

{\it Methods --}\label{meth} To model  $N_{p}$ strongly interacting three-colour (component/species) fermions trapped in an $L$-site ring-shaped lattice pierced by an effective magnetic flux $\phi$, we employ the SU($3$) Hubbard model
\begin{equation}\label{eq:Ham}
\mathcal{H} = \sum\limits_{j=1}^{L}\sum\limits_{\alpha=1}^{3}\bigg[-t( e^{\imath\frac{2\pi\phi}{L}}c_{j,\alpha}^{\dagger}c_{j+1,\alpha}+\textrm{h.c.})+\sum_{\beta>\alpha}U_{\alpha\beta}n_{j,\alpha}n_{j,\beta}\bigg]
\end{equation}
where $c_{j,\alpha}^{\dagger}$ creates a fermion with colour $\alpha$ on site $j$, and $n_{j,\alpha} = c^{\dagger}_{j,\alpha}c_{j,\alpha}$ is the local particle number operator. The parameters $t$ and $U_{\alpha\beta}$ denote the hopping amplitude and interaction strength respectively. We consider attractive interactions i.e. $U_{\alpha\beta}\!<\!0$ and $t=1$ fixes the energy scale. The effective magnetic field is realised through Peierls' substitution $t\rightarrow te^{\imath \frac{2\pi\phi}{L}}$. 
 
In  the  continuous limit of  vanishing lattice spacing or, equivalently, the dilute lattice limit $\nu=N_{p}/L\!\ll\!1 $ the physics of the system can be captured by the Gaudin-Yang-Sutherland model~\cite{sutherland1968further,takahashi1970many-body}.
According to the general Bethe Ansatz machinery, the energy of the system is obtained after a set of coupled non-linear equations - the Bethe equations - are solved for  the quasimomenta $k_{j}$ of the system. The spectrum is  obtained by such $k_{j}$, and results to be labelled by a specific set of quantum numbers~\cite{takahashibook,sutherland1968further}. 
Bound states result from complex values of the quasimomenta (see Supplementary). In the limit of large $UL/t\!\gg\! 1$, the spectrum of the SU($3$) Gaudin-Yang-Sutherland regime of the Hubbard model is obtained by solving a set of three equations, the so-called Takahashi equations, parametrized by $n_{1}$, $n_{2}$ and $n_{3}$ denoting the number of unpaired, pairs, and trions respectively (see Supplementary). We point out that the strongly attractive regime of  model~\eqref{eq:Ham} cannot be recast into a Lai-Sutherland anti-ferromagnet since the condition of one particle per site cannot be achieved~\cite{sutherland1975model,capponi2016phases}.  
Bound states of different nature can arise in systems described by model~\eqref{eq:Ham}: CSF bound states, wherein two colours form a bound pair with the other colour remaining unpaired; trions, wherein all the three colours form a bound state. CSF bound states can be achieved by breaking the SU(3) symmetry~\cite{catelani2008phase,Batchelor2010exact,kuhn2012phase,Klingschat2010exact,Guan2013Fermi}. Here, we break the SU(3) symmetry explicitly in the canonical ensemble by  choosing asymmetric interactions between the different colours (see \cite{Batchelor2010exact,kuhn2012phase,Guan2013Fermi} for symmetry breaking in the grand-canonical ensemble by adjusting the chemical potentials for each species). In the following, $U$ refers to symmetric interactions between all colours.
\begin{figure}[h!]
    \centering
    \includegraphics[width=0.85\linewidth]{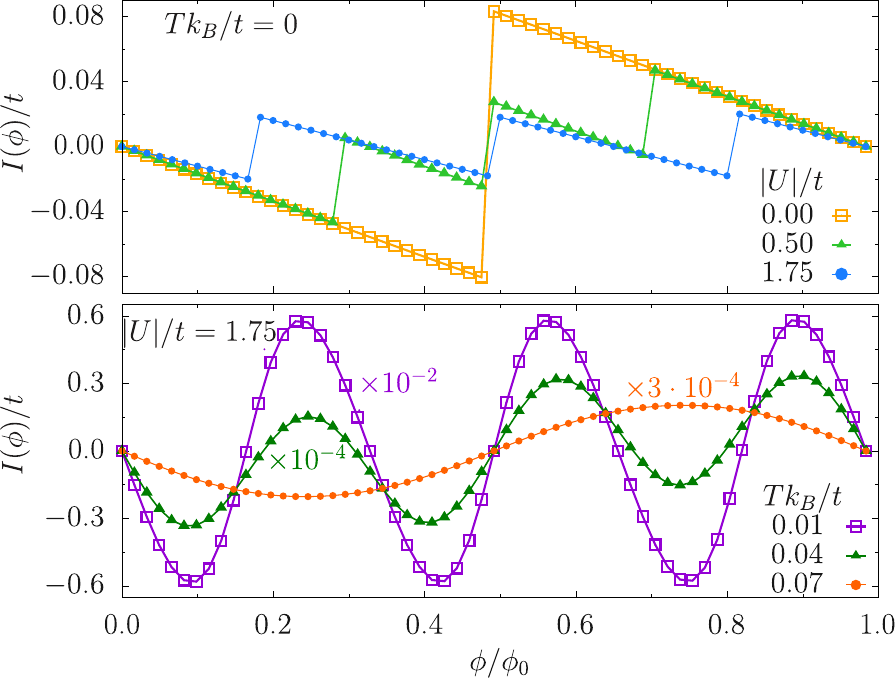}%
    \put(-17,150){(\textbf{a})}
    \put(-17,77){(\textbf{b})}
    \caption{Persistent current $I(\phi)/t$ of SU(3) fermions for various interactions $U/t$ (temperatures $Tk_{B}/t$) in the upper (lower) panel. In (\textbf{a}) for $Tk_{B}/t=0$, the persistent current fractionalizes with increasing $U/t$. The bare period $\phi_{0}/t$ is reduced to $\phi_0/N$ for large $U/t$. For fixed $U/t$ in (\textbf{b}), the persistent current regains the period $\phi_0$ upon increasing $Tk_{B}/t$. The results were obtained by exact diagonalization with $N_{p} =3$, $L=15$. The lines are meant to be a guide to the eye for the reader, to aid in perceiving the fractionalization with increasing interaction.} 
    \label{fig:tempcurr}
\end{figure}

SU(3) bound states of model~\eqref{eq:Ham} have been recently studied through correlation functions due to their relevance in emulating quark matter~\cite{Capponi2008molecular,Klingschat2010exact,pohlmann2013trion}. It is important to stress such systems are only analogues as they lack key features of quark matter such as string breaking and colour charge screening. The probe we use to study the system is the persistent current $I(\phi )$, which is the response to the effective magnetic flux threading the system: $I(\phi) = -\partial F(\phi)/\partial\phi$, with $F$ being the 
system's free energy in the canonical ensemble~\cite{byers1961theoretical}. The zero temperature persistent current arises only from the ground-state energy $E_{0}$, such that  $I(\phi) = -\partial E_{0}(\phi)/\partial\phi$. 
Relying on the experimental capability of addressing fermions of  different colours separately~\cite{Sonderhouse2020}, especially to analyse the broken SU($3$) cases, 
we  utilise the species-wise persistent current: $I_\alpha=-\partial F_\alpha(\phi)/\partial\phi$. We point out that this calculation cannot be easily implemented in Bethe ansatz. The reason being that the species-wise persistent current has to be done with two-point correlations, which is very challenging in Bethe ansatz.
\begin{figure*}[t]
    \centering
    \includegraphics[width=\linewidth]{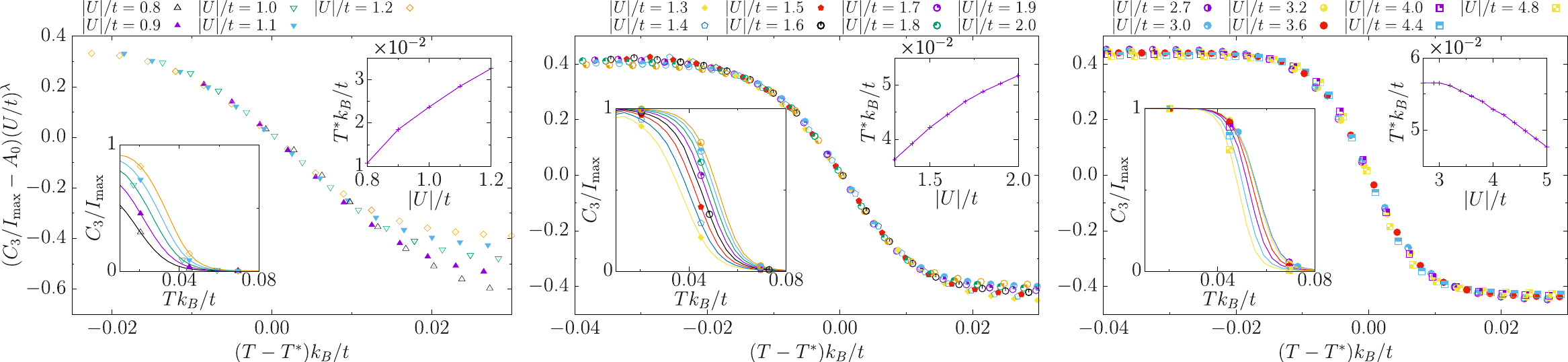}%
    \put(-485,18){(\textbf{a})}
    \put(-322,18){(\textbf{b})}
    \put(-150,18){(\textbf{c})}
    \caption{Interplay between temperature $Tk_{B}/t$ and interaction $U/t$ for the persistent current $I(\phi)/t$, by monitoring the Fourier weight $C_{3}/t$ of the current that reflects its tri-partite periodicity. This coefficient decreases upon increasing temperature showing the breakdown of trions and in turn deconfinement. The three plots demonstrate that $C_3(T^*)/I_{max}(T^*)$  obeys distinct laws in the different regimes of interaction discussed in the text, where $I_{max}/t$ is the maximum persistent current used to re-scale the Fourier weight across different interactions. The constant shift in $C_3(T^*)/I_{max}(T^*)$, is fixed for all curves by $A_{0}=0.5$. Top right insets display the temperature displacement $T^{*}k_{B}/t$ as a function of $U/t$. Lower left insets show the raw data as a function of $Tk_{B}/t$.  $T^{*}k_{B}/t$ is defined by $C_3 (T^*)/I_{max}(T^*)=1/2$. In the regime of weak $U/t$ displayed in (\textbf{a}),  $\lambda=-1.25$ and $T^{*}k_{B}/t$ is an increasing function of $|U|/t$. For intermediate $U/t$ depicted in (\textbf{b}) $\lambda=-0.33$ and $T^{*}k_{B}/t$ is still increasing, but with a different algebraic law. For the strong $U/t$ regime in (\textbf{c}), $\lambda=-0.1$ and $T^{*}k_{B}/t$ is decreasing with $|U|/t$. All results were obtained with exact diagonalization for $N_{p}=3$, $L=15$ with the temperature $Tk_{B}/t$ ranging from 0.01 to 0.08. Note that in the lower left insets only a few data points are displayed to enhance the readability of the figure.}
    \label{fig:scales}
\end{figure*} 

An important  result in the field due to Leggett, states that the energy of a many-body system displays periodic oscillations with the  flux $\phi$~\cite{leggett1991}. 
If single particle states are involved in the persistent current, the period is the bare one $\phi_{0}=\hbar/mR^2$, with $m$ and $R$ denoting the atoms' mass and ring radius, respectively. When bound states are formed, the corresponding effective mass leads to a 
reduced periodicity in $\phi_{0}$~\cite{byers1961theoretical,polo2020exact,pecci2021probing}.

In cold atom systems, it has been demonstrated that most features of $I(\phi)$ can be observed through time-of-flight (TOF) imaging~\cite{amico2005quantum}. TOF expansion entails the calculation of the particle density pattern 
$    n_{\alpha}(\textbf{k}) = |w(\textbf{k})|^{2}\sum_{j,l}e^{\imath \textbf{k}\cdot(\textbf{x}_{j}-\textbf{x}_{l})}\langle c_{j,\alpha}^{\dagger}c_{l,\alpha}\rangle $
where $w(\textbf{k})$ are the Fourier transforms of the Wannier function and $\textbf{x}_{j}$ denotes the position of the lattice sites in the plane of the ring. From the experimental side, $n(\mathbf{k})$ can be accessed through contrast image measurement of the density distribution after a  TOF expansion of the condensate is carried out by switching off the confinement potential~\cite{amico2005quantum,amico2021atomtronic}. The variance of the width of the momentum distribution is given by
$\sigma^{(\alpha)}_{n_k}=\sqrt{\langle \hat{n}_\alpha^2\rangle-\langle \hat{n}_\alpha\rangle^2}$.
We utilise a combination of numerical methods such as exact diagonalization and DMRG~\cite{whitedmrg,itensor}, as well as Bethe ansatz results whenever possible, in order to identify and characterize the bound states of SU($3$) fermions, for systems with 
an equal number of particles $N_{p}$ per colour.  In particular, zero temperature properties are addressed through DMRG; finite temperature results are obtained through exact diagonalization. As we already pointed out, the model is Bethe ansatz integrable in the case of certain parameters and filling fractions. Such constraints make the finite temperature analysis out of reach of Bethe ansatz.

\begin{figure*}[t]
    \centering
    \includegraphics[width=\linewidth]{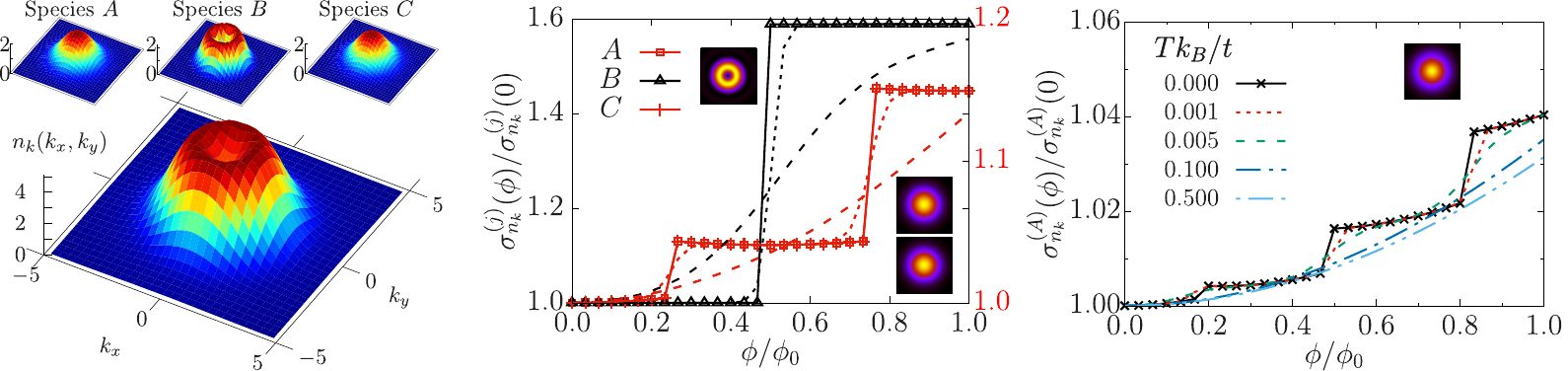}%
    \put(-520,112){(\textbf{a})}
    \put(-355,112){(\textbf{b})}
    \put(-175,112){(\textbf{c})}
    \caption{%
    (\textbf{a}) TOF expansion of the CSF configuration. Main (top) panel displays the TOF expansion, $n(k)$, for all (each) colours. Interactions are set to $|U_{AB}|/t=|U_{BC}|/t=0.01$ and $|U_{AC}|/t=5$.
    (\textbf{b}-\textbf{c})Variance of width of the TOF expansion, $\sigma_{n_k}(\phi)$, against the effective magnetic flux $\phi$. Panel (\textbf{b}) shows the CSF configuration and $|U_{AB}|/t=|U_{BC}|/t=0.01$ and $|U_{AC}|/t=5$ for $T k_{B}/t= \{0,0.01,0.1\}$, solid, dotted and dashed lines respectively. Panel (\textbf{c}) shows the trionic configuration at $|U| /t= 5$ for different $T k_{B}/t$. Insets next to the curves in (\textbf{b}) show the momentum distribution $n_k(k_x,k_y)$ at $\phi=1$ of each component while in (\textbf{c}) we only show one colour due to SU(3) symmetry. The presented results are done for $N_{p}=3$, $L=10$ using exact diagonalization.}
    \label{fig:TOF}
\end{figure*}

{\it Zero temperature persistent current of SU(3) bound states --}\label{pcp}
 For small $U$, $I(\phi)$ is found to be a function with a period of the bare flux quantum $\phi_{0}$.  
 However, for stronger $U$, $I(\phi)$ displays  fractionalization reducing its period. In contrast to attracting bosons~\cite{naldesi2019rise,polo2020exact,polo2020quantum} or repulsing $N$-component fermions~\cite{yu1992persistent,chetcuti2020persistent}, for attracting $N$-components fermions with symmetric interactions,
 the reduction of the period is dictated by $N$ irrespective of $N_{p}$. 
 In the  SU($3$) symmetric case, we find that trions are  formed for arbitrary small attraction (see \cite{pohlmann2013trion,capponi2016phases}) in the three particle sector. This is corroborated by exact results based on the  Bethe Ansatz analysis of the Gaudin-Yang Sutherland model (see Supplementary) and by the analysis of three-body correlation functions. In this regime and  large $UL/t$, the analysis based on  Takahashi's equations, demonstrates a perfect tri-partition of the period $\phi_0/3$ that amounts to the formation of a three-body bound state --Fig.~\ref{fig:main} (\textbf{b}). 
 By Bethe ansatz analysis, we find the exact expression of the zero temperature $I(\phi)$ in the continuous limit for a system consisting solely of trions in the limit of large $U$:
\begin{equation}\label{eq:currbethe}
I(\phi) = -6\bigg(\frac{2\pi}{L}\bigg)^{2}\bigg[\frac{K_{a}}{3}+\phi\bigg]
\end{equation}
where $K_{a}$ denotes the aforementioned quantum numbers. 
This implies that as the flux is increased, $K_{a}$ need to shift to counteract this increase in flux (see Supplementary). Here, level crossings occur between the ground and excited
states, causing the fractionalization. Discrete excitations can only partially compensate for the increase in flux, causing oscillations with a reduced period of $1/N$,
thereby accounting for the ‘size’ of the bound state.  For the CSF, which is out of reach of Bethe ansatz due to SU(3)-symmetry breaking, $I(\phi)$ displays a halved periodicity $\phi_{0}/2$ for the paired colours and a bare periodicity $\phi_{0}$ for the unpaired colour --Fig.~\ref{fig:main} (\textbf{a}). Further confirmation on the nature of the bound states is achieved through the analysis of correlation functions (see Supplementary).

{\it Finite temperature effects and colour deconfinement --} 
The interplay between temperature $T$ and {\it{attractive}} interaction $U$ leads to remarkable effects in the  persistent current $I(\phi)$: Besides the generic smoothening of the saw-tooth behaviour, finite temperature leads to specific changes in the frequencies of $I(\phi)$ depending on the interaction -Fig.~\ref{fig:tempcurr}. Such a phenomenon is consistent with the thermal effects on two-component fermions with repulsive interactions~\cite{patu2021temperature}, which we show also holds for attractive interactions, irrespective of the number of particles in the system (see Supplementary). 
In the following, we discuss the thermal effects on the SU(3) symmetric interaction case (finite temperature CSF case is discussed in Supplementary). For small and moderate interactions,  the analysis shows that $I(\phi)$, and its frequency in particular,  arise from  thermal fluctuations populating the scattering states (for the band structure of the system see~\cite{pohlmann2013trion} and Supplementary).  
Here, the relevant parameter is the relative size between interaction $U$ and thermal fluctuations $T$ (measured in units of $t/k_B$). At moderate $U$ the bound states can remain well-defined for large $U/T$, whilst  for smaller values of  $U/T$ the bound states' deconfinement occurs because the temperature makes scattering states accessible.
On increasing $U$, the relevant contributions to $I(\phi)$ come from the bound states' sub-band only. For such a `gas of bound states', the periodicity of $I(\phi)$ changes because  the  temperature allows the different frequencies  of the excited states to contribute to the current.  In this regime, since the level spacing between the bound states energy levels decreases,  the thermal effects  are increasingly relevant by increasing $U$.        

To study the specific dependence of $I(\phi)$ on $T$ and $U$, we analyse its power spectrum.  Specifically, we consider the Fourier weight $C_3$ of $I(\phi)$  corresponding to trion formation at different $U$ values and follow its decay with increasing temperature (see Appendix for the explicit definition). Such a weight corresponds to the reduced tri-partite periodicity corresponding to the formation of trions. This coefficient is rescaled by the maximum amplitude $|I_{max}|$ of $I(\phi)$. We find that $C_3/|I_{max}|=U^{-\lambda} G({T-T^*})$, in which  $\lambda$, $T^*$ (see Fig.~\ref{fig:scales} caption) and the function $G$ as shown in  Fig.~\ref{fig:scales} (\textbf{a}) and (\textbf{c}) are markedly distinct in the aforementioned different interaction regimes (see also Supplementary). Remarkably, in the cross-over region between the colour deconfinement region and the bound states gas, the two regimes result to be indistinguishable --Fig.~\ref{fig:scales} (\textbf{b}).    

{\it Time of flight --}\label{tof} Persistent currents can be probed in cold atom systems through TOF. To read out the nature of the states in our system, it is important that such images arise as an interference pattern of the gas wave functions. For the specific case of  coherent neutral matter circulating with a given angular momentum, a characteristic hole is displayed. Due to the reduced coherence, no holes have been found in TOF for bound states  (see exponential decay of correlations in the Supplementary)~\cite{naldesi2020enhancing,pecci2021probing,pecci2021phase,chetcuti2022interference}. Nevertheless, current states and the corresponding angular momentum quantization emerge in the variance $\sigma$ of the width of the momentum distribution as discrete steps. Here, trions display three steps in $\sigma$ reflecting the reduced tri-partite periodicity of the current. For CSF states, we find a characteristic TOF  with decreased density in the center of the interference pattern.  By analysing, the different colour contributions to the TOF, we deduced that the images arise as superpositions of the hole corresponding to the delocalised weakly coupled species and the smeared peak corresponding to the bound state of the paired particles. Such bound states are found to be  characterized in TOF by just two steps in $\sigma$ (reflecting the particle pairing) --Fig.~\ref{fig:TOF}. 

{\it Conclusion --}\label{conc}
In this paper, we studied the bound states of attracting three-component fermions through the frequency of the persistent current $I(\phi)$ both at zero and finite temperature. To this end, we apply a combination of Bethe ansatz and numerical methods that, especially for the finite temperature results, are among the very few non-perturbative approaches that can be applied to our system.  Our analysis hinges on the fact that the  {\it{effective}} flux quantum, defined by the frequency of $I(\phi)$ provides information on the nature of the particles involved in $I(\phi)$~\cite{leggett1991,byers1961theoretical,naldesi2019rise,chetcuti2020persistent}.  For our specific system of attractive SU(3) fermions, 
such a frequency  {\it  indicates that three-colour bound states are formed, irrespective of the number of particles}. This $N=3$ case is the general feature we find for SU($N$) {\it attracting} fermions whose bound states are formed by $N$ particles; in contrast to {\it repulsive} fermions and attractive bosons in which the frequency is fixed by the number of particles. 
Our analysis can clearly distinguish between trions and CSFs: the first are characterized by the persistent currents of the three species displaying a periodicity that is increasingly reduced by interaction until reaching $1/3$ of the original periodicity (for large interaction); CSFs, instead, result in persistent currents having two different periodicities for the different species --Fig.~\ref{fig:main}.

Finite temperature $T$ induces specific changes in the persistent current frequency. We analysed the interplay between interaction and thermal fluctuations quantitatively and {obtained specific laws describing it}. For  mild interactions, the frequency of $I(\phi)$ changes as result of the population of the scattering states.  Indeed, we observe that the phenomenon occurs as a  {\it crossover  from a colourless bound state to coloured multiplets, governed by the  ratio $U/T$ without an explicit $SU(3)$ breaking} --Fig.~\ref{fig:tempcurr} and Fig.~\ref{fig:scales} (\textbf{a}). Although specific non-perturbative effects near the QCD transition are missed by our analogue system (such as string breaking and colour charge screening), the  bound states' deconfinement in this regime displays similarities  with  the Quark-Gluon plasma formation at large $T$ and small baryonic density~\cite{satz2013probing}. Moreover, the introduction of a chemical potential, $\mu$, could permit us to study of the ``critical line" in the $T$-$\mu$ plane, in analogy with QCD at finite temperature and density, related with relativistic, but lower energy, heavy ion collisions and with the equation of state in the neutron stars core~\cite{Andronic2018,Annala2020}. However this aspect requires a dedicated forthcoming analysis.
For stronger attraction, the system defines a gas of bound states separated from the scattering states by a finite energy gap.  {\it In this regime,  a `single particle'   thermal persistent current arises from the  combination of the frequencies characterizing the different energy levels  in the bound state sub-band} --Fig.~\ref{fig:scales} (\textbf{c}). 
In the cross-over region between the scattering-states dominated  and the gas of bound states, the change of the frequency of $I(\phi)$  takes place with an identical functional dependence on $U$ and $T$ --Fig.~\ref{fig:scales}. 
For increasing interaction, the bound states' sub-band gets tighter and, therefore the temperature is increasingly  relevant to wash out the fractionalization of the persistent current's periodicity.  

The suggested implementation of our work is provided by cold atoms. Thus, we studied the time-of-flight images of the system obtained by releasing the cold atoms from the trap --Fig.~\ref{fig:TOF}.  Additionally, we point out that recent advancements in the platform of programmable tweezers have paved the way to experimentally realize fermionic ring lattices~\cite{tweezer}. This way, we believe that most of the presented results can be tested experimentally within the current cold atoms quantum technology infrastructure.

To conclude, we briefly comment on the phenomenon of three-body losses. In principle, these losses can be  due to the presence of Efimov states~\cite{naidon2017efimov}. However, we note that, unless specifically tuned, Efimov states occur as excited states~\cite{williams2009evidence}, and as such they are not expected to impact the nature of the ground-state. Although, the formation of a stable three-component Fermi gas has been experimentally realized~\cite{ottenstein},  experimental studies of three-body losses in one-dimensional SU(3) fermions are still lacking. In the case of spinless and  SU(2) fermions, three-body losses are suppressed on account of the Pauli exclusion principle. 
However, the exclusion statistics of SU(3) fermions could allow the possibility of three-body recombination. 
In the case of bosonic systems, this phenomenon is suppressed in one-dimensional systems and thus by employing a similar logic, we expect a similar behavior occurs in our system~\cite{williams2009evidence,blochsuppression,naidon2017efimov}. Whilst there is not a general consensus on the explanation, the theoretical analysis on bosonic systems indicates that the characteristic increase in the 1D scattering length~\cite{olshanii1998atomic} may result in a lower probability of forming resonant bound states~\cite{mehta2007three}.

\begin{acknowledgments}\label{ack}
{\it Acknowledgments --}
We acknowledge fruitful discussions with G.~Catelani, G.~Marchegiani, G.~Sierra, F.~Scazza and T.~Haug.
\end{acknowledgments}

\bibliography{ref.bib}

\setcounter{equation}{0}

\onecolumngrid
\newpage

\begin{appendix}

\section{Supplementary Material}

\noindent In the following, we provide supporting details of the results found in the manuscript \textit{Probe for bound states of SU(3) fermions and colour deconfinement}.

\subsection{Mapping from the SU(\textit{N}) Hubbard model to the Gaudin-Yang-Sutherland model}

\noindent Here, we present the derivation of the Gaudin-Yang-Sutherland model 
from the SU($N$) Hubbard model as the continuous limit 
of vanishing lattice spacing.\\

\noindent The density of fermions in the lattice denoted by $D$ can be defined as $D = N_{p}/(L\Delta )$ with $N_{p}$ being the number of particles, $L$ is the number of sites and $\Delta$ stands for the lattice spacing. The filling factor $\nu$ is related to the lattice spacing as
\begin{equation}\label{eq:hh1}
\nu = \frac{N_{p}}{L} = \frac{N_{p}}{\Delta L} \Delta
\end{equation}
Therefore, in the continuous limit of vanishing lattice spacing $\Delta\rightarrow 0$ and finite particle density $\frac{N_{p}}{\Delta L}$, also the filling must be  accordingly small. 
For the anticommutation relations to hold in the continuous limit, the fermionic operators have to be rescaled as
\begin{equation}\label{eq:hh2}
c_{i,\alpha}^{\dagger} = \sqrt{\Delta}\Psi^{\dagger}_{\alpha}(x_{i}) \hspace{20mm} n_{i,\alpha} = \Delta\Psi^{\dagger}_{\alpha}(x_{i})\Psi_{\alpha}(x_{i})\hspace{10mm}\mathrm{where}\hspace{2mm}x_{i} = i\Delta
\end{equation}
where $\Psi^{\dagger}$ is the creation field operator obeying the standard anti-commutation relations $\{\Psi_{\alpha}(x),\Psi^{\dagger}_{\beta}(y)\} = \delta_{\alpha,\beta}\delta (x-y)$ and $\{\Psi^{\dagger}_{\alpha}(x),\Psi^{\dagger}_{\beta}(y)\} = 0$. \\

\noindent Utilizing Equation~\eqref{eq:hh2}, 
the SU($N$) Hubbard model is mapped onto the Fermi gas quantum field theory in the following way
\begin{equation}\label{eq:hh3}
\mathcal{H}_{\mathrm{SU}(N)} = t\Delta^{2}\mathcal{H}_{FG}-2N_{p}   
\end{equation}
\begin{equation}\mathcal{H}_{FG} = \int \bigg[(\partial_{x}\Psi^{\dagger}_{\alpha})(\partial_{x}\Psi_{\alpha}) + c\sum\limits_{\alpha <\beta}^{N}\Psi^{\dagger}_{\alpha}\Psi^{\dagger}_{\beta}\Psi_{\beta}\Psi_{\alpha}\bigg]\end{equation}
with $\alpha$ and $\beta$ denoting different SU(N) colours,  and $c = \frac{U}{t\Delta}$. The Fermi gas field theory is the quantum field theory of the Gaudin-Yang-Sutherland model. 
This means that the eigenstates of $\mathcal{H}_{FG}$ as $\ket{\psi (\lambda )} = \sum\limits_{\alpha_{1} \hdots\alpha_{N_{p}}}^{N} \int \chi (\textbf{x}|\lambda )\Psi^{\dagger}_{\alpha_{1}} (x_{1})\hdots\Psi^{\dagger}_{\alpha_{N_{p}} }(x_{N_{p}})\ket{0}\textrm{d}\textbf{x}$ 
are provided by 
$\chi (\textbf{x}|\lambda )$ being eigenfunctions of the Gaudin-Yang-Sutherland Hamiltonian
\begin{equation}\label{eq:hh4}
  \mathcal{H}_{GYS} = -\sum\limits_{\alpha=1}^{N}\sum\limits_{i=1}^{N_{\alpha}} \frac{\partial^{2}}{\partial x_{i,\alpha}^{2}} + 4c\sum\limits_{i<j}\sum\limits_{\alpha,\beta}\delta (x_{i,\alpha} - x_{j,\beta})
\end{equation}

\noindent An important property of the continuous limit is that the centre of mass dynamics separate from the relative coordinates. 
This connection is lost in the lattice theories~\cite{naldesi2020enhancing}.
This is one of the key features explaining the lack of integrability of the lattice regularization of continuous theories. A heuristic way to understand why the system becomes integrable in the continuous limit (for both attractive and repulsive interactions), is to  note that the diluteness condition makes the  probability of more than two particles interacting vanish~\cite{frahm1995on}. Therefore, the system gets `non-diffractive' in the sense that the scattering matrix obeys the Yang-Baxter relations~\cite{sutherland2004beautiful}. For repulsive interactions,  the SU($N$) Hubbard model is integrable also for large $U$ and one species per site (the physics of the system is captured by the Lai-Sutherland model).
However, the latter condition cannot be met for attractive interactions. Hence, the SU($N\!>\!2$) Hubbard model with {\it attractive} interactions is not Bethe Ansatz integrable in the Lai-Sutherland regime. \\

\noindent Despite the formal mapping between the lattice  SU($N$) and  continuous theories works for arbitrary interaction, the attractive case of negative $U$ requires extra care. The catch lies in the formation of bound states that can have a correlation length comparable or smaller than $\Delta$. 
In particular, for $\Delta$ larger than the coherence length, one will not be able to observe the correlation functions' decay as the latter is overshadowed by the former.
This is particularly evident for small-sized systems. The condition outlined above can be satisfied for large-sized systems and small values of the interaction.  In order to obtain a meaningful continuous limit to be quantitatively comparable with the lattice theory at small $\nu$, the  vanishing lattice spacing $\Delta$ should come with a suitable rescaling of $U$. While such an aspect has been analysed for bosonic theories (see f.i.~\cite{oelkers2007ground,naldesi2019rise}), for fermions it still requires investigation. Even though one needs to rescale $U$ for SU(2) fermions, this problem can be circumvented due to the Lieb-Wu Bethe ansatz, which allows one to validate the Bethe ansatz results with numerical methods. However, the SU($N\!>\!2$) Hubbard model is not Bethe ansatz integrable for lattice systems. As such, one is not able to compare numerical results with the exact solution.  Therefore, particular care  when using the Bethe ansatz for SU($N$) is required. Nonetheless, the continuous limit is Bethe ansatz integrable by the Gaudin-Yang-Sutherland model and it is vital in understanding the underlying physics of the model.

\subsection{Bethe ansatz equations for the Gaudin-Yang-Sutherland model.}

\noindent The Gaudin-Yang-Sutherland model describes a system of interacting fermions with SU($N$) symmetry residing on a ring of size $L_{R}$ threaded by an effective magnetic flux $\phi$ as follows~\cite{sutherland1968further,decamp2016high}
\begin{equation}\label{eq:app1}
\mathcal{H}= -\sum\limits^{N}_{l=1}\sum\limits^{N_{l}}_{j=1}\bigg(-\imath\frac{\partial}{\partial x_{j,l}}-\frac{2\pi\phi}{L_R}\bigg)^{2}+4c\sum\limits_{j,k}\sum\limits_{l<n}^{N}\delta (x_{j,l}-x_{k,n})
\end{equation}
where $c$ is the interaction, $N_{l}$ is the number of particles with colour $l$ of SU($N$) symmetry with $l=1\hdots N$. The model is integrable by Bethe ansatz and is given by the following set of equations.
\begin{equation}\label{eq:app2}
e^{\imath (k_{j}L_{R}-\Phi)} = \prod\limits_{\alpha =1}^{M_{1}}\frac{k_{j}-\lambda_{\alpha}^{(1)}+\imath c}{k_{j}-\lambda_{\alpha}^{(1)}-\imath c}  \hspace{4mm} j=1,\hdots ,N_{p} 
\end{equation}

\begin{equation}\label{eq:app3}
\prod_{\substack{\beta = 1 \\\beta\neq\alpha}}^{M_{r}} \frac{\lambda_{\alpha}^{(r)} - \lambda_{\beta}^{(r)} +2\imath c}{\lambda_{\alpha}^{(r)} - \lambda_{\beta}^{(r)} -2\imath c}  = \prod\limits_{\beta = 1}^{M_{r-1}} \frac{\lambda_{\alpha}^{(r)} - \lambda_{\beta}^{(r-1)} +\imath c}{\lambda_{\alpha}^{(r)} - \lambda_{\beta}^{(r-1)} -\imath c}\cdot \prod\limits_{\beta = 1}^{M_{r+1}} \frac{\lambda_{\alpha}^{(r)} - \lambda_{\beta}^{(r+1)} +\imath c}{\lambda_{\alpha}^{(r)} - \lambda_{\beta}^{(r+1)} -\imath c}\hspace{4mm}\alpha = 1,\hdots , M_{r}
\end{equation}
for $r=1,\hdots ,N-1$, where $M_{0}=N_{p}$ and $\lambda_{\beta}^{0} =k_{\beta}$. $N_{p}$ denotes the number of particles, $M_{r}$ corresponds to the number of particles in a given colour, with $k_{j}$ and $\lambda_{\alpha}^{(r)}$ being the charge and spin rapidity respectively, with $\Phi = 2\pi\phi$. The energy corresponding to the state for every solution of these equations is $E = \sum\limits_{j=1}^{N_{p}}k_{j}^{2}$. 

\noindent For SU(3) fermions, one obtains a set consisting of three coupled transcendental equations. 
\begin{equation}\label{eq:app4}
e^{\imath (k_{j}L_{R}-\Phi)} = \prod\limits_{\alpha=1}^{M_{1}}\frac{k_{j}-\Lambda_{\alpha} +\imath c}{k_{j}-\Lambda_{\alpha}-\imath c}  \hspace{4mm} j=1,\hdots ,N_{p} 
\end{equation}

\begin{equation}\label{eq:app5}
\prod_{\substack{\beta = 1 \\\beta\neq\alpha}}^{M_{1}} \frac{\Lambda_{\alpha} - \Lambda_{\beta} +2\imath c}{\Lambda_{\alpha} - \Lambda_{\beta} -2\imath c}  = \prod\limits_{\beta = 1}^{N_{p}} \frac{\Lambda_{\alpha} - k_{\beta}+\imath c}{\Lambda_{\alpha} - k_{\beta} -\imath c}\cdot \prod\limits_{\beta = 1}^{M_{2}} \frac{\Lambda_{\alpha} - \lambda_{\beta} +\imath c}{\Lambda_{\alpha} - \lambda_{\beta} -\imath c} \hspace{4mm}\alpha = 1,\hdots , M_{1}
\end{equation}

\begin{equation}\label{eq:app6}
\prod_{\substack{b=1\\ b\neq a}}^{M_{2}} \frac{\lambda_{a} - \lambda_{b} +2\imath c}{\lambda_{a} - \lambda_{b} -2\imath c}  =  \prod\limits_{\beta = 1}^{M_{1}} \frac{\lambda_{a} - \Lambda_{\beta} +\imath c}{\lambda_{a} - \Lambda_{\beta} -\imath c} \hspace{4mm} a = 1,\hdots , M_{2}
\end{equation}
where $\lambda^{(1)}$ and $\lambda^{(2)}$ were changed to $\Lambda$ and $\lambda$ respectively, for the sake of convenience. These equations are valid for both repulsive and attractive interactions. The difference between the two regimes arises from the fact that in the attractive regime, the Bethe ansatz solutions are complex strings for the quasimomenta $k_{j}$ in the ground-state, due to the formation of bound states. In the case of SU($N$) fermions with strongly attractive interactions $L_R|c|\!\gg\!1$, the quasimomenta $k_{j}$ may appear as a bound state composed of $m$ particles, with the length (number) of particles ranging from 2 to $N$, where $N$ is 3 in the case considered. The real part in the strings of charge and spin rapidities is given by $\lambda^{(m-1)}$. As such, it will accompany all the strings. Unpaired fermions have real quasimomenta $k_{j}$.  \\

\begin{figure}[h!]
    \includegraphics[width=0.7\textwidth]{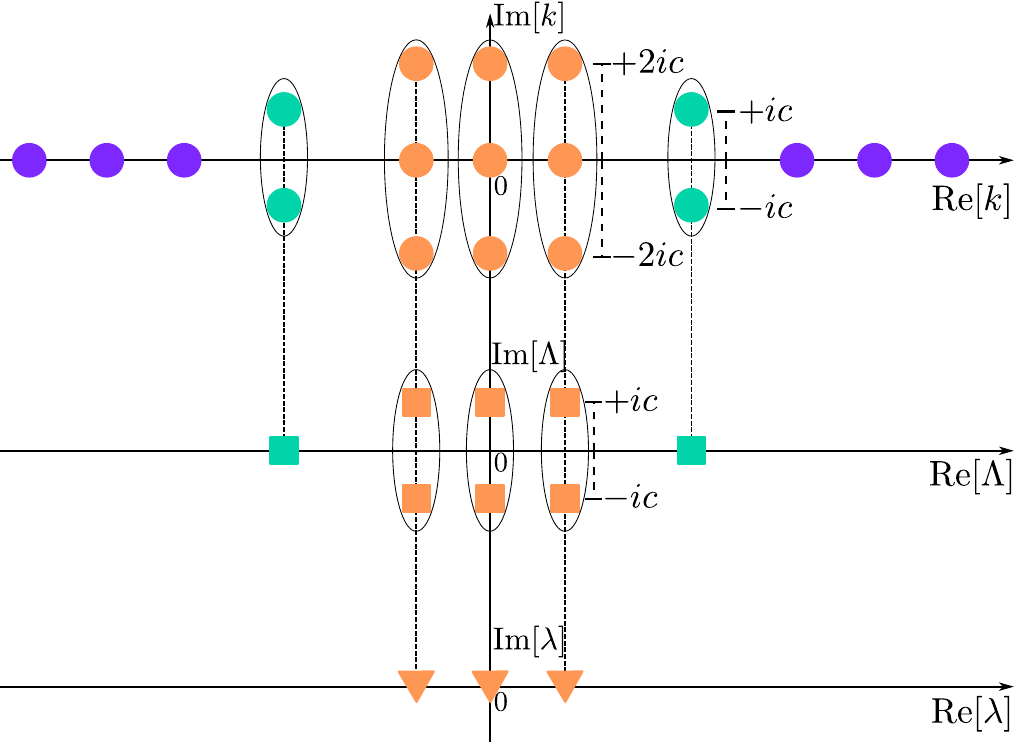}%
    \caption{Figure of merit for the quasimomenta string configuration in the ground-state for $N=3$, with $n_{1}=6$ (purple), $n_{2} = 2$ (green) and $n_{3} = 3$ (orange). The bound states in the system are represented by an oval structure, and they are present both for the charge (circles) and spin rapidities (squares and triangles). Accompanying a charge bound state of length $m$, are spin bound states of decreasing length, with $m=2$ being the minimum length one needs to form a bound state. The real part of the charge and spin bound states of length $m$ is given by the spin rapidity corresponding to $\lambda^{(m-1)}$. The figure above is based on the one in~\cite{takahashi1970many-body}.}
    \label{fig:taka}
\end{figure}

\noindent Considering the SU(3) case, we can have $n_{1}$ unpaired fermions ($m=1$), $n_{2}$ pairs of two-body bound states ($m=2$) and $n_{3}$ three-body bound states ($m=3$), with $N=n_{1}+2n_{2}+3n_{3}$, $M_{1}=n_{2}+2n_{3}$ and $M_{2}=n_{3}$. Consequently, the quasimomenta and corresponding spin strings are as follows~\cite{Guan2013Fermi}
\begin{align}
\centering
\label{eq:app9}
    & k^{1}_{\alpha} = k_{\alpha} &
    &k^{2}_{\beta}=  
\begin{cases}
    \Lambda_{\beta}+\imath c\\
    \Lambda_{\beta}-\imath c             
\end{cases}  &
&k^{3}_{a}= 
\begin{cases}
    \lambda_{a}+2\imath c\\
    \lambda_{a} \\
    \lambda_{a}-2\imath c             
\end{cases}& \\
     &\color{white}0 & 
     &\Lambda_{\beta} = \Lambda_{\beta} &
     &\Lambda_{a}= 
\begin{cases}
    \lambda_{a}+\imath c\\
    \lambda_{a}-\imath c             
\end{cases}& \\
    &\color{white}0 &
    &\color{white}0&
    &\lambda_{a}=\lambda_{a}&
\end{align}
for each case respectively where $\alpha = 1,\hdots ,n_{1}$, $\beta = 1\hdots, n_{2}$ and $a = 1,\hdots, n_{3}$.  It is important to note that the strings given above are ideal strings and they are only applicable in the limit $L_R|c|\!\gg\!1$. If one does not satisfy this condition, then one needs to include the term $\mathcal{O}(\imath\delta |c|)$ to every string, with $\delta$ being a very small quantity whose order is that of $\exp (-L_R|c|)$~\cite{yin2011effective}. For $L_R|c|\!\gg\! 1$, $\mathcal{O}(\imath\delta |c|)$ goes to zero and one is able to utilise the ideal strings. In the following, we will always assume that we are in this limit unless explicitly stated. \\

\noindent Substituting the ideal strings into the Bethe ansatz equations~\eqref{eq:app4} to~\eqref{eq:app6}, we obtain the following three equations.
\begin{equation}\label{eq:app12}
k_{j}L_R = 2\pi I_{j} - 2\sum\limits_{\alpha = 1}^{n_{2}}\arctan \bigg(\frac{k_{j}-\Lambda_{\alpha}}{c}\bigg) - 2\sum\limits_{a = 1}^{n_{3}}\arctan \bigg(\frac{k_{j}-\lambda_{a}}{2c} \bigg) +\Phi  
\end{equation}
\begin{align}\label{eq:app13}
 2\Lambda_{\alpha}L_R = 2\pi J_{\alpha} &-2\sum\limits_{j = 1}^{n_{1}}\arctan \bigg(\frac{\Lambda_{\alpha}-k_{j}}{c}\bigg) - 2\sum\limits_{\substack{\beta = 1 \\ \beta\neq\alpha}}^{n_{2}}\arctan \bigg(\frac{\Lambda_{\alpha}-\Lambda_{\beta}}{2c}\bigg) - 2\sum\limits_{a = 1}^{n_{3}} \arctan \bigg(\frac{\Lambda_{\alpha}-\lambda_{a}}{c}\bigg) \\
 &- 2\sum\limits_{a = 1}^{n_{3}}\arctan \bigg(\frac{\Lambda_{\alpha}-\lambda_{a}}{3c}\bigg) +2\Phi \nonumber
\end{align}

\begin{align}\label{eq:app14}
3\lambda_{a}L_R = 2\pi K_{a} &-2\sum\limits_{j = 1}^{n_{1}}\arctan \bigg(\frac{\lambda_{a}-k_{j}}{2c}\bigg)-  2\sum\limits_{\alpha = 1}^{n_{2}}\arctan \bigg(\frac{\lambda_{a}-\Lambda_{\alpha}}{c}\bigg)-  2\sum\limits_{\alpha = 1}^{n_{2}}\arctan \bigg(\frac{\lambda_{a}-\Lambda_{\alpha}}{3c}\bigg) \\
&-2\sum\limits_{\substack{b = 1 \\ b\neq a}}^{n_{3}}\arctan \bigg(\frac{\lambda_{a}-\lambda_{b}}{2c}\bigg)-2\sum\limits_{\substack{b = 1 \\ b\neq a}}^{n_{3}} \arctan \bigg(\frac{\lambda_{a}-\lambda_{b}}{4c}\bigg)+3\Phi \nonumber
\end{align}
for $j=1,\hdots, n_{1}$, $\alpha = 1,\hdots , n_{2}$ and $a = 1,\hdots , n_{3}$ with $I_{j}$, $J_{\alpha}$ and $K_{a}$ being the quantum numbers associated to the charge, first and second spin rapidities respectively. These are called Takahashi's equations for SU(3) fermions with attractive delta interaction~\cite{takahashi1970many-body}. The total energy is given by
\begin{equation}\label{eq:app15}
E  = \sum\limits_{j=1}^{n_{1}}k_{j}^{2} + \sum\limits_{\alpha=1}^{n_{2}}(2\Lambda_{\alpha}^{2}-2c^{2}) + \sum\limits_{a=1}^{n_{3}}(3\lambda_{a}^{2}-8c^{2})
\end{equation}
and it is clear that for any value of the interaction $c$, the energy of a trion will always be lower than that of a CSF, which in turn is lower than that of an unpaired particle.  Consequently, this implies that in the ideal string limit, one will have trions whenever possible. As a result, in order to form an unpaired particle or a CSF in the continuous limit for $L_R|c|\!\gg\!1$, the SU(3) symmetry needs to be broken. \\

\begin{figure*}[h!]
    \centering
    \includegraphics[width=0.6\linewidth]{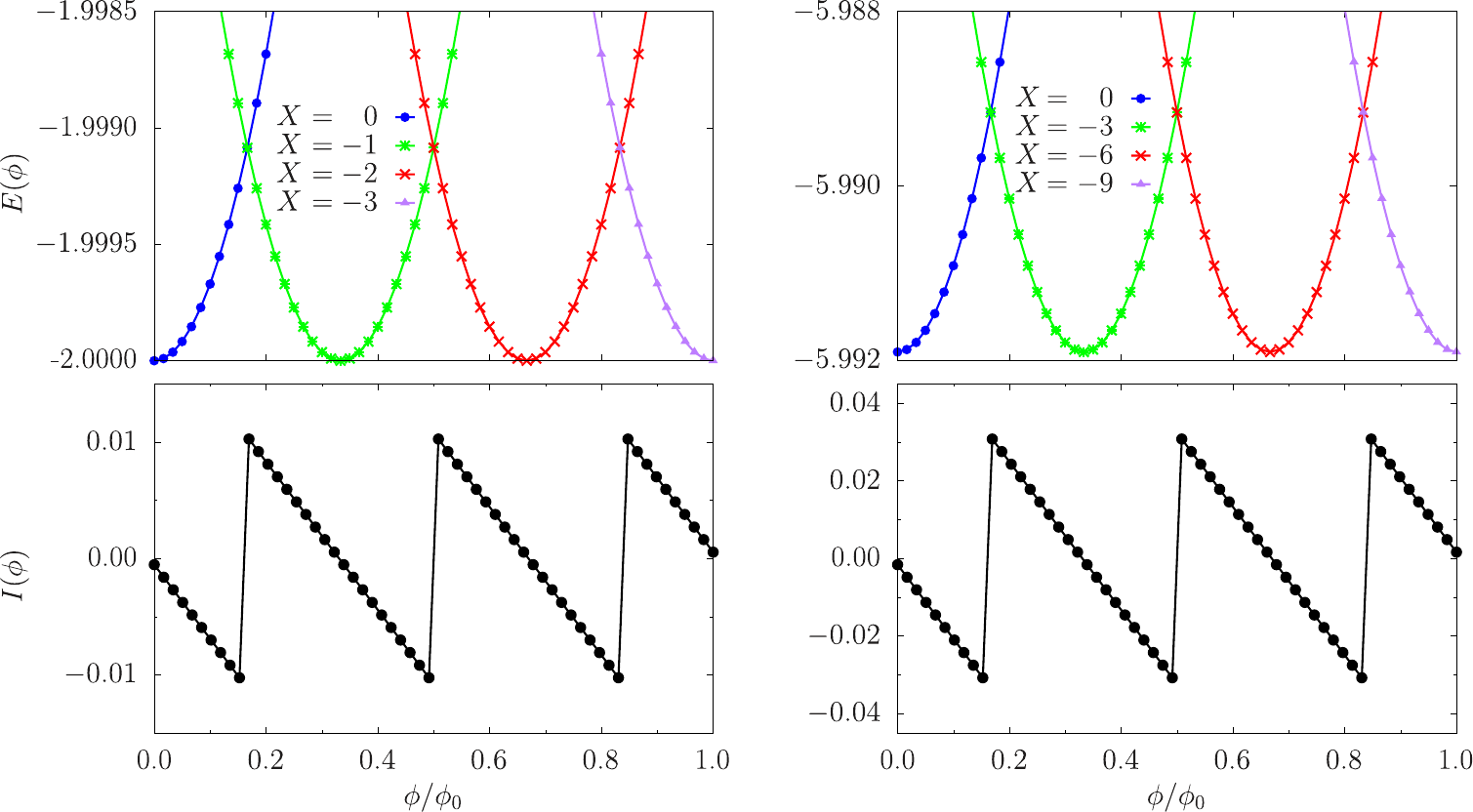}%
     \put(-172,150){(\textbf{a})}
     \put(-19,150){(\textbf{b})}
    \put(-172,80){(\textbf{c})}
     \put(-19,80){(\textbf{d})}
    \caption{The top row (\textbf{a}) and (\textbf{b}) show how the Bethe ansatz energies $E(\phi/\phi_{0})$ need to be characterized by the quantum numbers ($X = \sum_{a}K_{a}$) in order to have the actual ground-state for increasing flux $\phi/\phi_{0}$.  The bottom row (\textbf{c}) and (\textbf{d}) depicts the corresponding persistent current $I(\phi/\phi_{0})$ for a system consisting solely of trions with $N_{p}=3,9$ respectively..
   All the presented results were obtained with the Takahashi equations of the Gaudin-Yang-Sutherland model for $L_R=60$ and $|c|=2$. }
    \label{fig:bet}
\end{figure*}
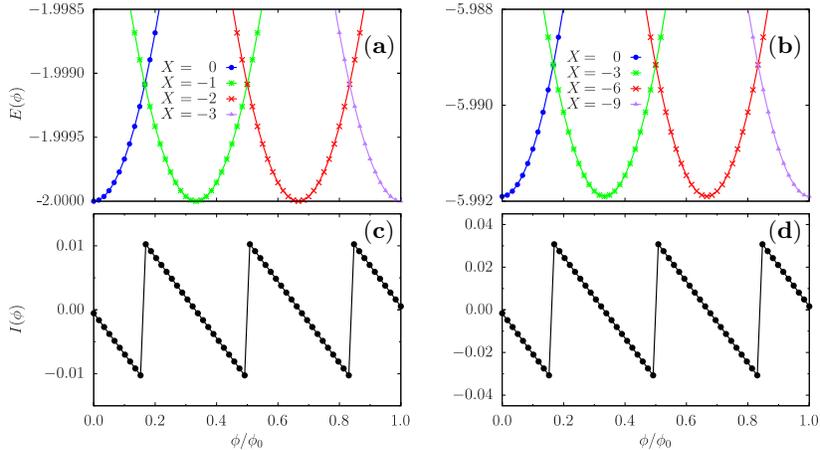

\noindent In the limit of $|c|\rightarrow\infty$, the Bethe ansatz equations for $n_{1}=n_{2}=0$ read as 
\begin{equation}\label{eq:inf1}
\lambda_{a} = \frac{2\pi}{L_R}\bigg(\frac{1}{3}K_{a} + \phi\bigg)
\end{equation}
Substituting Equation~\eqref{eq:inf1} into Equation~\eqref{eq:app15}, 
\begin{equation}\label{eq:inf2}
E = 3\bigg(\frac{2\pi}{L_R}\bigg)^{2} \sum\limits_{a=1}^{n_{3}}\bigg[\bigg(\frac{K_{a}}{3}\bigg)^{2} + \phi^{2} + 2\frac{K_{a}}{3}\phi  \bigg] -8n_{3}c^{2}
\end{equation}
At zero temperature, the persistent current of the system is defined as $I(\phi) = -\partial E_{0}/\partial\phi$. Consequently, the persistent current as $|c|\rightarrow\infty$ becomes
\begin{equation}\label{eq:inf3}
 I(\phi) = -6\bigg(\frac{2\pi}{L_R}\bigg)^{2}\sum\limits_{a}^{n_{3}}\bigg[\frac{K_{a}}{3}+\phi\bigg] 
\end{equation}
which resembles the equation that is obtained for SU($N$) fermions with repulsive interactions as $c\rightarrow\infty$~\cite{yu1992persistent,chetcuti2020persistent}, but with the important difference that here the period of the persistent current is reduced by the number of colours  $N$ instead of by the number of particles $N_p$.   From Equation~\eqref{eq:app15}, we know that in order to reduce the energy we require that $\lambda\rightarrow 0$. Therefore, when the flux $\phi$, which is a positive quantity, is increased, we require that $K_{a}$ takes on a negative value so as to counteract the increase in flux. Changing the quantum numbers creates excitations in the ground-state by having level energy crossings between the ground and excited states. These excitations are quantized in nature and so they can only compensate the increase in flux partially, which in turn causes periodic oscillations with a reduced period of $1/3$ that in this case corresponds to the number of particles in the bound state (trions).\\
\begin{table}[ht!]
\centering
\begin{tabular}{|c|c|}
    \hline
     Magnetic Flux & $K_{a}$ \\
     \hline
     0.0 - 0.1 &  \{-1,0,1\} \\
     0.2 - 0.5 &  \{-2,-1,0\} \\
     0.6 - 0.8 & \{-3,-2,-1\} \\
     0.9 - 1.0 & \{-4,-3,-2\} \\
     \hline
\end{tabular}
     \caption{Quantum number configurations with the flux for SU(3) fermions with $N_{p}=9$ for a system containing only 3 trions. }\label{tab:t}
\end{table}

\noindent The  different mechanism  leading to the fractionalization for the repulsive and the attractive regimes can be highlighted by studying the Bethe ansatz equation as $|c|\rightarrow\infty$. Remarkably, for a system containing only trions, the Bethe ansatz equations are decoupled, in that the rapidity $\lambda_{a}$ is only dependent on its own quantum number $K_{a}$. However, for the repulsive case this does not occur. Indeed, a similar expression as the one in Equation~\eqref{eq:inf1} is obtained for repulsive interactions, but the latter is still dependent on the spin quantum numbers of the other particles~\cite{chetcuti2020persistent}. Therefore, whilst in the repulsive case a single spinon excitation acts to counteract the flux for the whole system, this is not the case for attractive interactions. Indeed, the decoupling between the ``different'' bound states (trions), implies that all the trions corresponding to a rapidity $\lambda_{a}$, need to shift their quantum number $K_{a}$ to counteract the increase in flux and minimize the energy.  An example on how to change the quantum numbers with increasing flux is presented in Table~\eqref{tab:t}. 

\subsection{Solution of the Bethe ansatz equations}

\noindent In this section, we  will highlight the nature of the bound states in the constraints of small $L_R|c|$. For the case $N_p=3$, we will prove that trions are formed for any small value of the interaction.\\

\begin{figure}[ht!]
    \includegraphics[width=0.33\textwidth]{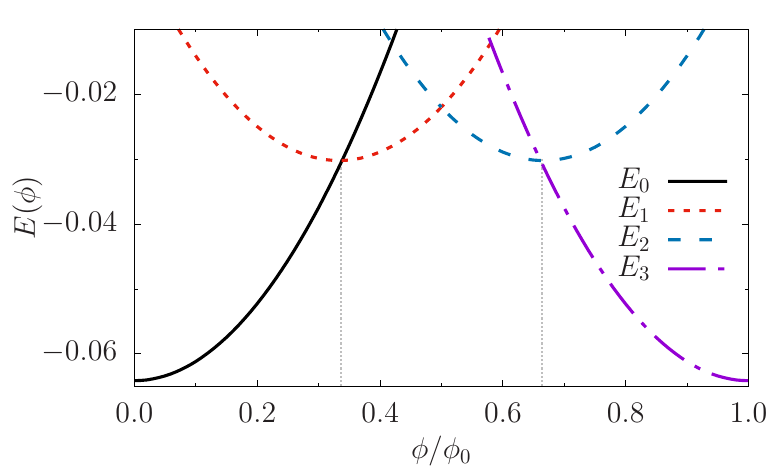}%
    \includegraphics[width=0.33\textwidth]{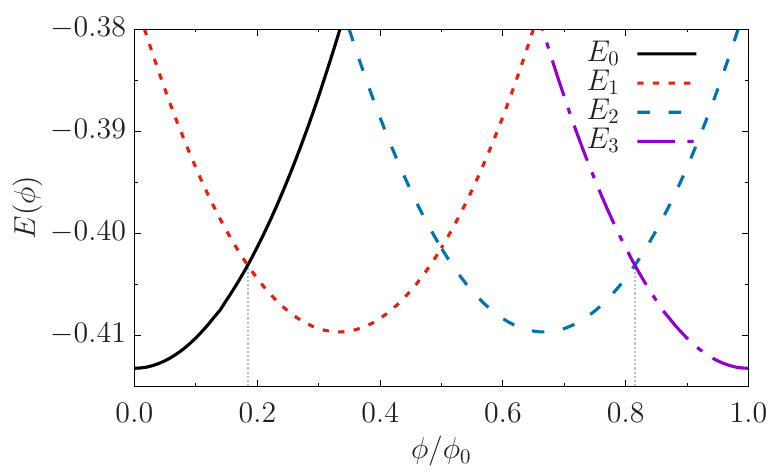}%
    \includegraphics[width=0.33\textwidth]{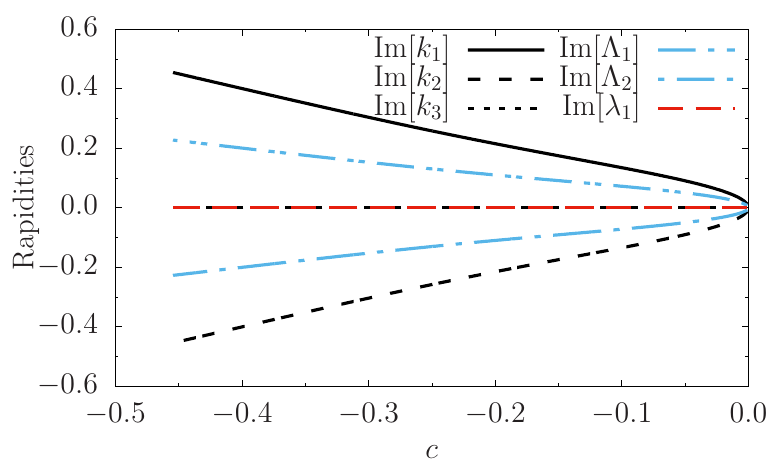}
    \put(-475,35){(\textbf{a})}
    \put(-306,35){(\textbf{b})}
    \put(-140,35){(\textbf{c})}
    \caption{Results obtained from solving the Bethe ansatz equations in the product form for $N_p=3$ and $N=3$. Two interactions are considered: (\textbf{a}) $|c|=0.155$ where it just starts fractionalizing  (\textbf{b}) $|c|=0.45$ where the $N$-times fractionalization is almost reached. The size in the Bethe equations is fixed to $L_R=20$ for all curves. (\textbf{c}) The complex part of the rapidities against interaction. As the interaction increases, the complex part of the rapidities goes to the ideal string limit as outlined in Equations~\eqref{eq:app9}. Note that the real part of the rapidities is always zero for all $k_j;\Lambda_\alpha;\lambda_\beta$ for this case. }
    \label{fig:prodbethe}
\end{figure}

\noindent As we discussed above, the main technical feature of the attracting fermions is that the ground-state is formed by bound states. In integrable theories, this corresponds to complex Bethe rapidities.  In the previous section, we sketched how such complex solutions, for  $L_R|c|\!\gg\! 1$, are arranged in the Takahashi string solutions.  For small interactions and small systems,  Takahashi's equations assuming ideal string configurations of the solution, cannot be used to solve the Bethe ansatz equations and access the energy. Below, we provide the details of the solution of the Bethe ansatz equations, i.e. the product form Equations~\eqref{eq:app3} to~\eqref{eq:app6}, without any constraints on $c$ and $L_R$.  We solve these coupled equations numerically, following an iterative approach in which one starts close to the $c\lesssim 0$ and iteratively looks for solutions that are continuous in the energy and do not display jumps in the logarithmic form. \\

\noindent Figs.~\ref{fig:prodbethe} (\textbf{a}) and (\textbf{b}) show the Bethe ansatz solution for two different interactions. As a deviation of the string hypothesis, we observe that initially the full $N$-times periodicity is not reached (thin dotted lines) until the interactions are strong enough. Nonetheless, even at interactions of $|c|\approx 0.5$, we already observe some of the $N$-times periodicity of the energy (here seen through the crossing points between the parabolas, now occurring near $1/2N$).
However, by looking at the rapidities Fig.~\ref{fig:prodbethe} (\textbf{c}), we can confirm that, for the three particle case considered here, a trion appears for any negative interactions. Particularly, both $k_j$ and $\Lambda_\alpha$ are complex for any $c<0$.

\subsection{Correlations}
\noindent In this section, we study the three-body correlation function defined as:
\begin{equation}\label{eq:three}
T_{i,j,k} = \langle c^{\dagger}_{i,A}c^{\dagger}_{j,B}c^{\dagger}_{k,C}c_{k,C}c_{j,B}c_{i,A}\rangle \;,
\end{equation}
where $i,j,k$ denote the lattice sites and $A,B,C$ are the colours and $c_{j\alpha}^{\dagger}$ denoting the typical fermionic creation operator for colour $\alpha$ at site $j$. Similar correlations function have been studied in~~\cite{Capponi2008molecular,Klingschat2010exact,pohlmann2013trion}. Through this correlation function, we analyse its decay for both trions and CSF to understand the nature of the bound states in our lattice system.

\begin{figure}[h!]
    \centering
    \includegraphics[width=0.495\linewidth]{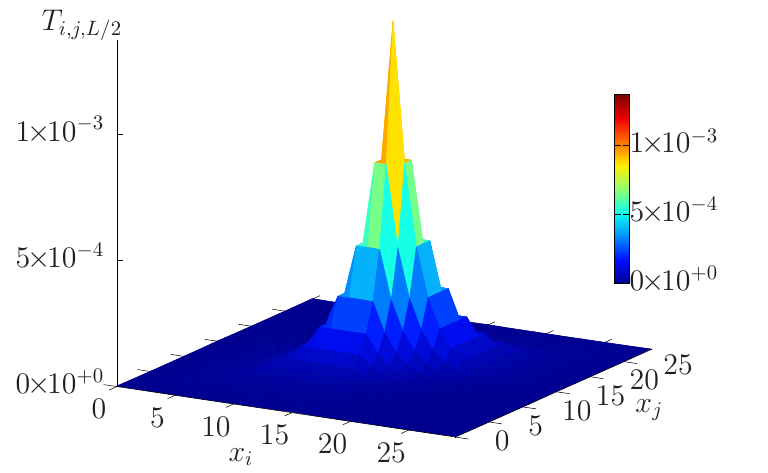}
    \includegraphics[width=0.495\linewidth]{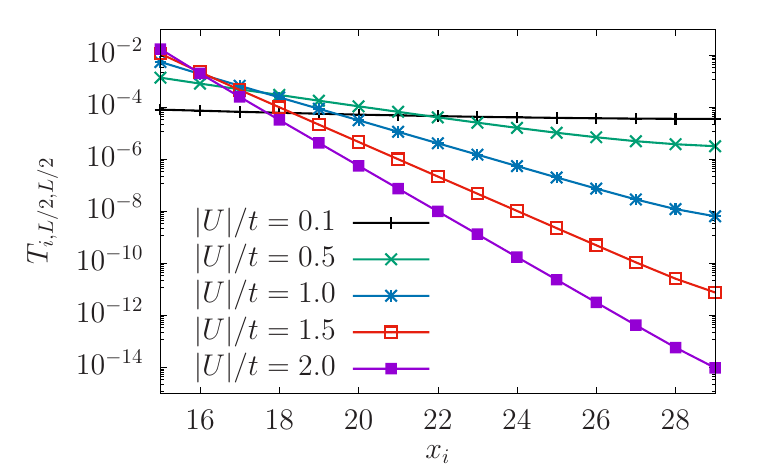} 
    \put(-510,144){(\textbf{a})}
    \put(-246,144){(\textbf{b})}
    \caption{(\textbf{a}) Three-body correlation function, $T_{i,j,L/2}$, and its decay (\textbf{b}) for $L = 30$ and $N_{p}$ = 3 with a trion configuration for $|U_{AB}|/t = |U_{AC}|/t = |U_{BC}|/t= 0.5$. All results were obtained with exact diagonalization for $\phi = 0$. }
    \label{fig:densdensdenstrion}
\end{figure}

\begin{figure}[h!]
\centering
    \includegraphics[width=0.495\linewidth]{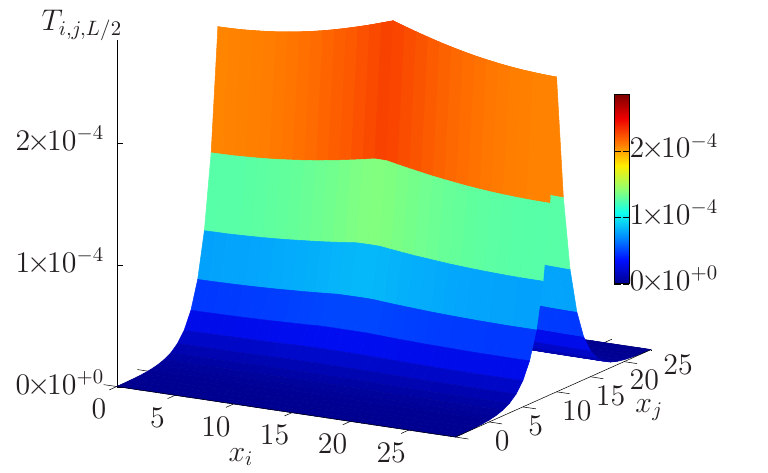}
    \includegraphics[width=0.495\linewidth]{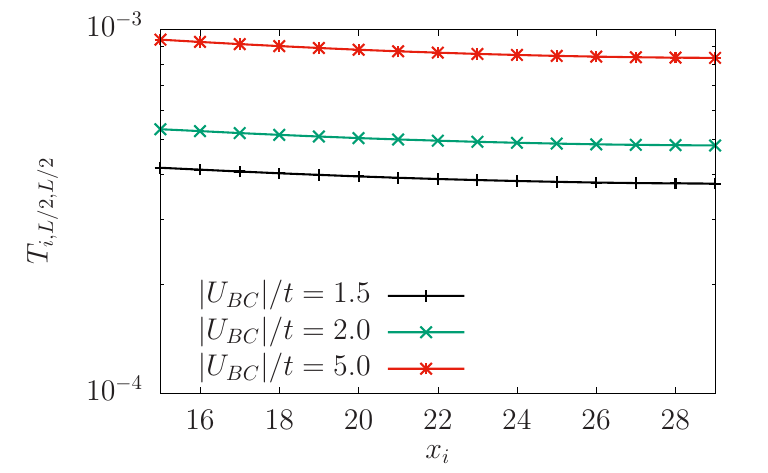}
    \put(-510,144){(\textbf{a})}
    \put(-246,144){(\textbf{b})}    
    \caption{
    (\textbf{a}) Three-body correlation function, $T_{i,j,L/2}$, for $L = 30$ and $N_{p}$ = 3 with a CSF configuration for $|U_{AB}|/t=|U_{AC}|/t= 0.01$ and $|U_{BC}|/t= 5$. (\textbf{b}) The decay of the correlation function for different values of $|U_{BC}|/t$ is depicted. All results were obtained with exact diagonalization for $\phi = 0$.}
        \label{fig:d3csf}
\end{figure}

\noindent Particularly, three-body correlations will display an exponential decay in $\{i,j,k\}$ whenever a three-body bound state exists. This should be contrasted with the results of such a correlator, $T_{i,j,L/2}$, for a CSF, which only displays an exponential decay on $j$, as the bound states are of a lower degree, i.e., formed by a bound state consisting of two particles. Here, we show two examples of such a correlator by fixing one of the axis, $T_{i,j,L/2}$, in Fig.~\ref{fig:densdensdenstrion} (\textbf{a}) and Fig.~\ref{fig:d3csf} (\textbf{a}). In addition, we also calculate, in their corresponding panels (\textbf{b}), their decay for different values of the interaction $U$. This decay shows us that the correlation length associated to the bound state decays exponentially for a trionic state, which means that the localization of the effective molecules increases. On the other hand, the CSF shows no such exponential decay. \\

\noindent Finally, we point out that demonstrating that the long-distance three-body correlation length dominates over any other correlator, gives us a clear signature of the nature of the bound state, and shows that in our exact diagonalization simulations (in the three particle sector), trions are formed as soon as attractive interactions are present in the system. These results agree with the Bethe ansatz findings shown in the previous sections of this Supplementary material.

\subsection{Symmetric vs asymmetric trions}

\noindent In this section, we investigate all the different ways that a trion can be formed by playing with the different interactions between the colours. \\
\begin{figure}[h!]
    \centering
    \includegraphics[width=0.9\linewidth]{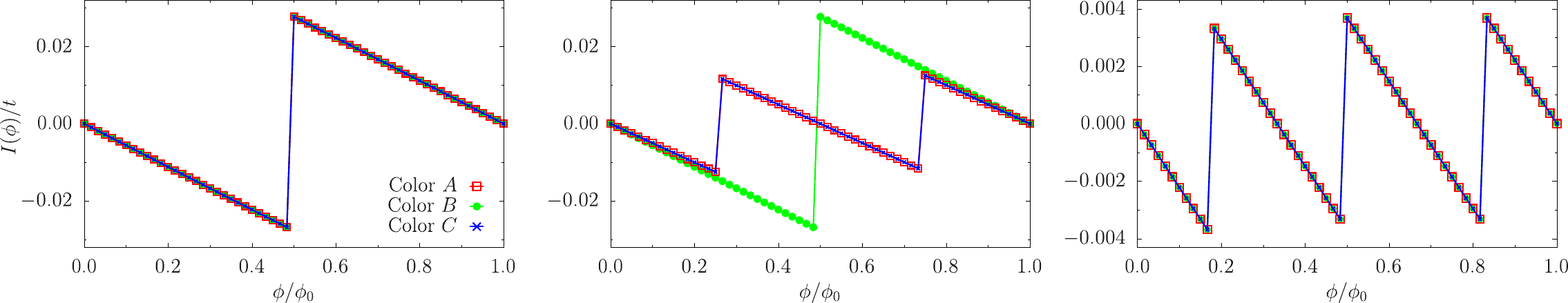}
    \put(-326,77){(\textbf{\textbf{a}})}
    \put(-171,77){(\textbf{\textbf{b}})}
    \put(-16,77){(\textbf{\textbf{c}})}
    \caption{Persistent current $I(\phi)/t$ against flux $\phi/\phi_{0}$ for the three main phases of SU(3) fermions: (\textbf{a}) unpaired, (\textbf{b}) csf and (\textbf{c}) trion. Results were obtained with exact diagonalization for $N_{p}=3$ and $L=15$. The lines are meant as a guide to the eye for the reader.}
    \label{fig:m1}
\end{figure}
\begin{figure}[h!]
    \centering
    \includegraphics[width=0.9\linewidth]{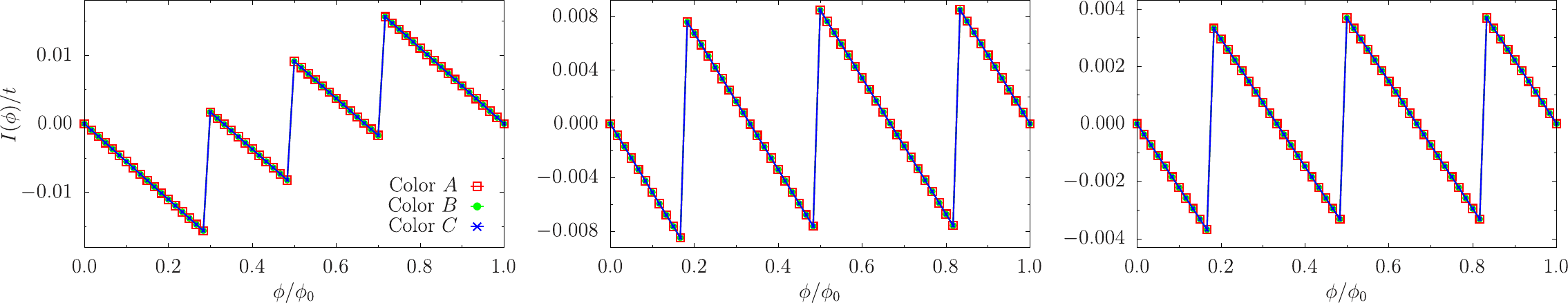}
    \put(-326,77){(\textbf{\textbf{a}})}
    \put(-171,77){(\textbf{\textbf{b}})}
    \put(-16,77){(\textbf{\textbf{c}})}
    \caption{Persistent current $I(\phi)/t$ against flux $\phi/\phi_{0}$ for the formation of a trion with symmetric interactions $|U|/t = |U_{AB}|/t=|U_{BC}|/t=|U_{AC}|/t$ for: (\textbf{a}) $|U|/t=0.5$, (\textbf{b}) $|U|/t=1$ and (\textbf{c}) $|U|/t=3$. Results were obtained with exact diagonalization for $N_{p}=3$ and $L=15$. The lines are meant as a guide to the eye for the reader.}
    \label{fig:m2}
\end{figure}

\noindent The first path to form a trion is displayed in Fig.~\ref{fig:m2} and it is achieved by using symmetric interactions, \textit{i.e.} equal interactions between the colours. Starting from a system with zero interactions like the one depicted in Fig.~\ref{fig:m1} (\textbf{a}), the persistent current fractionalizes with increasing interaction and experiences a periodicity change from $\phi_{0}$ to $\phi_{0}/N$ upon formation of the trion --Fig.~\ref{fig:m2} (\textbf{c}). \\

\begin{figure}[h!]
    \centering
    \includegraphics[width=0.9\linewidth]{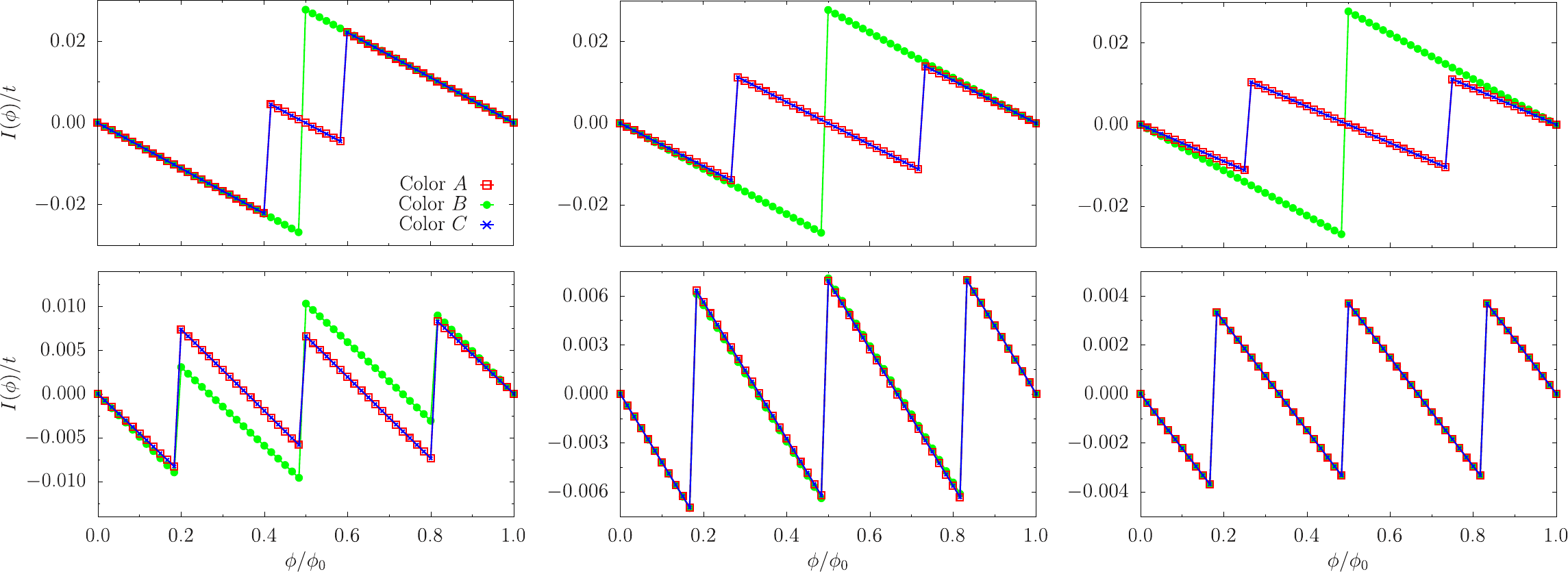}
    \put(-326,155){(\textbf{\textbf{a}})}
    \put(-170,155){(\textbf{\textbf{b}})}
    \put(-16,155){(\textbf{\textbf{c}})}
    \put(-326,77){(\textbf{\textbf{d}})}
    \put(-170,77){(\textbf{\textbf{e}})}
    \put(-16,77){(\textbf{\textbf{f}})}
    \caption{Persistent current $I(\phi)/t$ against flux $\phi/\phi_{0}$ for the formation of a trion by forming a CSF as an intermediate step. Starting from $|U_{AC}|/t=0.5$ in (\textbf{a}), one starts increasing the interactions between colours $A$ and $C$ until one forms a CSF in (\textbf{c}) with $|U_{AC}|/t=3$. The interactions $|U_{AB}|/t$ and $|U_{BC}|/t$ are kept at a value of 0.01. In the bottom panel, the interactions between colour $B$ to both colours $A$ and $C$ are increased from $|U_{AB}|/t=|U_{BC}|/t=0.5$ in (\textbf{d}) to $|U_{AB}|/t=|U_{BC}|/t=3$ in (\textbf{f}) at which a trion is  formed. Results were obtained with exact diagonalization for $N_{p}=3$ and $L=15$. The lines are meant as a guide to the eye for the reader.}
    \label{fig:m3}
\end{figure}

\noindent Another way to form a trion is by having an intermediate step and forming the other bound state of SU(3) fermions, which is a CSF. This path requires breaking the SU(3) symmetry by having an interaction between two colours, $|U_{AC}|$ in the case considered, being much larger than the ones between the other colours $|U_{AB}|$ and $|U_{BC}|$. As one increases the interaction $|U_{AC}|$, one can see the formation of the two-body bound state through the fractionalization of the persistent current --Fig.~\ref{fig:m3} (\textbf{a}) to (\textbf{c}).  Then, increasing the interactions $|U_{AB}|$ and $|U_{BC}|$ to match $|U_{AC}|$, causes further fractionalization and the persistent current obtains a reduced period of $\phi_{0}/N$. \\

\noindent Lastly, a trion can be formed by choosing asymmetric interactions between the colours. At variance with the formation of a CSF, we take two interactions $|U_{AB}|$ and $|U_{BC}|$ and make them significantly larger than $|U_{AC}|$. One observes that the persistent current undergoes fractionalization and eventually achieves the tri-partite periodicity. Interestingly enough, one has obtained the formation of a trion by breaking the SU(3) symmetry. However, this asymmetric trion can still be distinguished from an actual trion. The two distinguishing factors are the energy of an asymmetric trion being higher (less stable) than that of an actual trion, which is reflected in the persistent current, and the persistent current of the three species is not equal, thereby showing the SU(3) symmetry breaking. 

\begin{figure}[h!]
    \centering
    \includegraphics[width=0.85\linewidth]{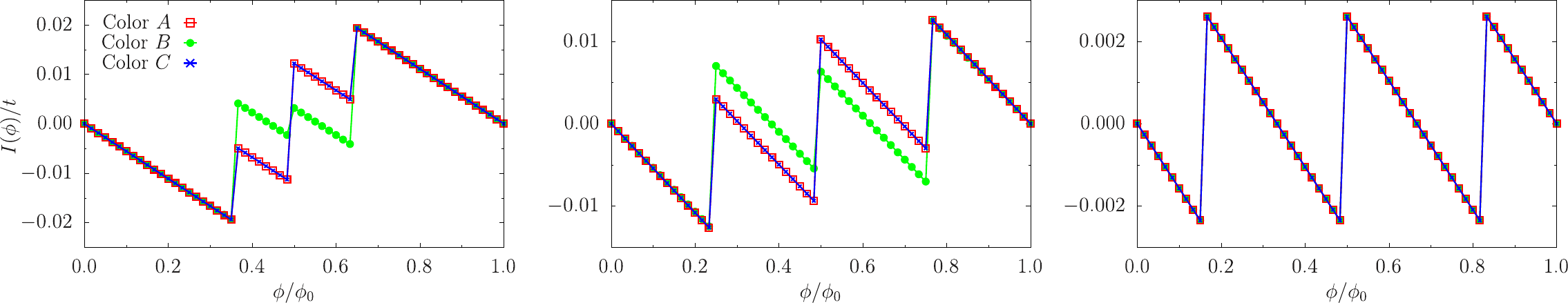}
    \put(-308,75){(\textbf{\textbf{a}})}
    \put(-163,75){(\textbf{\textbf{b}})}
    \put(-16,75){(\textbf{\textbf{c}})}
    \caption{Persistent current $I(\phi)/t$ against flux $\phi/\phi_{0}$ for the formation of a trion with asymmetric interactions. The interaction is increased between colours $A$ and $B$, and between $B$ and $C$ denoted by $|U_{AB}|/t$ and $|U_{BC}|/t$ respectively. At (\textbf{a}) the interactions $|U_{AB}|/t=|U_{BC}|/t=0.5$ are increased, until one forms a trion in (\textbf{c}) with $|U_{AB}|/t=|U_{BC}|/t=6$. The interaction $|U_{AC}|/t$ is kept at a constant value of 0.01 throughout. Results were obtained with exact diagonalization for $N_{p}=3$ and $L=15$. The lines are meant as a guide to the eye for the reader.}
    \label{fig:m4}
\end{figure}

\subsection{Parity effects}

\noindent The ground-state energy and in turn the persistent current of a system of spinless fermions is diamagnetic (paramagnetic) if the number of fermions is odd (even)~\cite{leggett1991}. In the case of an odd number of fermions, the ground-state energy of the system increases upon increasing the flux, thereby showing diamagnetic behaviour. On the other hand, for an even number of fermions, the ground-state energy decreases with increasing flux showing paramagnetic behaviour. This parity effect of the ground-state energy has been generalised to SU($N$) fermionic systems, whereby the current is diamagnetic (paramagnetic) for (2$n$+1)$N$ [(2n)$N$] fermions with $n$ being an integer number~\cite{chetcuti2020persistent}. It turns out that for strong repulsive interactions, the parity effect is washed out, with the system displaying diamagnetic behaviour irrespective of the number of fermions in the system. The reason being due the fractionalization that comes about from level crossings between the ground and excited states. The same holds true for SU(2) fermions with attractive interactions~\cite{pecci2021probing}. \textit{However, in the case of SU(3) fermions with attractive interactions, the parity effect persists even though one has fractionalization}. 
\begin{figure}[h!]
    \includegraphics[width=0.45\textwidth]{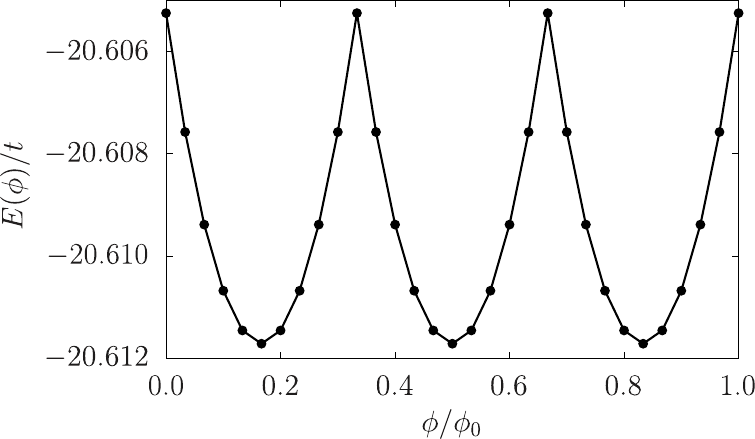}%
    \hspace{5mm}
    \includegraphics[width=0.45\textwidth]{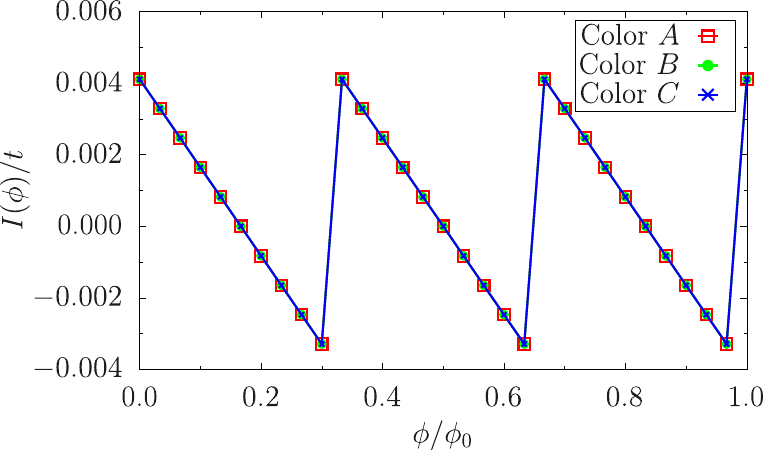}
    \put(-423,33){(\textbf{\textbf{a}})}
    \put(-185,33){(\textbf{\textbf{b}})}
    \caption{(\textbf{a}) Ground-state energy $E(\phi)/t$ and (\textbf{b}) the corresponding persistent current $I(\phi)/t$ as a function the flux $\phi/\phi_{0}$ for a system of $N_{p}=6$ with $L=20$ in a trion configuration for $|U_{AB}|/t = |U_{AC}|/t=|U_{BC}|/t=3$. All results were obtained with exact diagonalization. The lines in the right panel are meant as a guide to the eye for the reader.}
    \label{fig:partrion}
\end{figure}

\begin{figure}[h!]
    \includegraphics[width=0.45\textwidth]{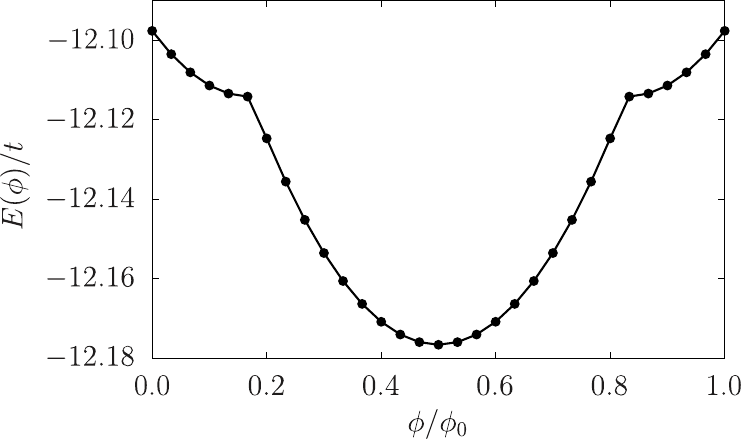}%
    \hspace{5mm}
    \includegraphics[width=0.45\textwidth]{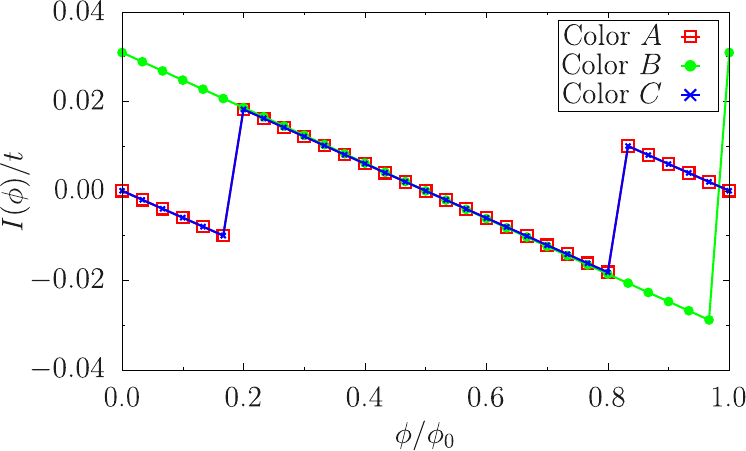}
    \put(-423,33){(\textbf{\textbf{a}})}
    \put(-185,33){(\textbf{\textbf{b}})}
    \caption{(\textbf{a}) Ground-state energy $E(\phi)/t$ and (\textbf{b}) the corresponding persistent current $I(\phi)/t$ as a function the flux $\phi/\phi_{0}$  for a system  of $N_{p}=6$ with $L=20$ in a CSF configuration for $|U_{AB}| /t= |U_{BC}|/t=0.01$ and $|U_{AC}|/t=1$. All results were obtained with exact diagonalization. The lines in the right panel are meant as a guide to the eye for the reader.}
    \label{fig:parcsf}
\end{figure}

\noindent Indeed, if we look at a system of six fermions with SU(3) symmetry with equal interactions, we find that as we increase the interaction, not only do we have fractionalization, but the system is still paramagnetic -- Fig.~\ref{fig:partrion}. This can be attributed to the fact that bound states of three fermions have an anti-symmetric wave function~\cite{yin2011effective}.  In contrast with the trion case, the system composed solely of CSFs has a different behaviour -- Fig.~\ref{fig:parcsf}. For the paired colours, the persistent current is diamagnetic~\cite{leggett1991,polo2020exact,pecci2021probing}. However, the remaining unpaired particles display diamagnetic (paramagnetic) behaviour if they are odd (even), indicating that they are nearly free.

\subsection{Finite temperature}

\begin{figure}[h!]
    \centering
    \includegraphics[width=0.45\linewidth]{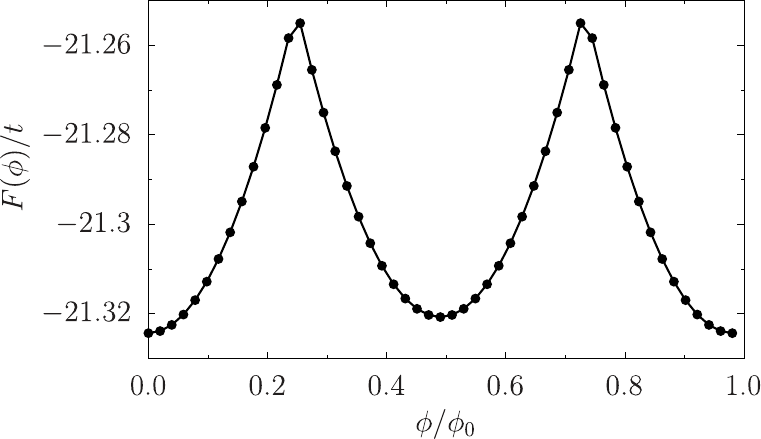}
    \includegraphics[width=0.45\linewidth]{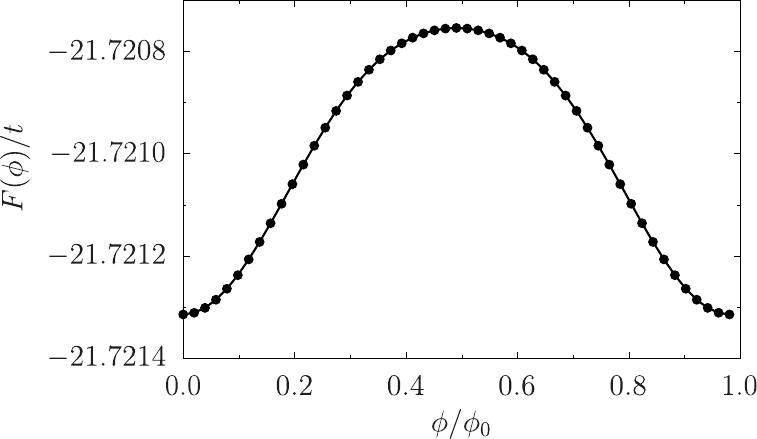}
    \caption{Free energy $F(\phi)/t$ as a function of the effective magnetic flux $\phi/\phi_{0}$ for $T k_{B}/t=0.0$ (left) and $T k_{B}/t=0.3$ (right). As the temperature increases, the energy loses its characteristic fractionalization and re-attains its single particle frequency. The results were obtained using exact diagonalization for $N_{p}=6$ fermions with SU(2) symmetry with $L=10$ and $|U|/t=6$.}
    \label{fig:SU(2)temp}
\end{figure}

\noindent Recently for two-component fermions with repulsive interactions~\cite{patu2021temperature}, it was shown how thermal effects affect the periodicity of the persistent current. In this paper, we show that a similar effect occurs for SU(2) fermions with attractive interactions (see Fig.~\eqref{fig:SU(2)temp}) and extended this result for SU(3) attracting fermions in the main text. \\

\noindent Here, we lay out the methodology that was employed to investigate the finite temperature effects on the persistent current, as well as its dependence on the interplay between temperature and interaction. This was done by performing the Fourier series on the finite temperature persistent current $I_{T}$.

\begin{equation}\label{eq:four1}
   C_{n} = \frac{1}{P}\int\limits_{P}X(x)\cdot e^{-\frac{2\imath\pi}{P} n x}\textrm{d}x
\end{equation}
where $C_{n}$ denotes the Fourier coefficient, $P$ is the period of the function and $X$ is the periodic function, which in our case is $I_{T}$. The persistent current at finite temperature is calculated via the current operator in the following manner: 
\begin{equation}\label{eq:current1}
     I_{T}(\phi/\phi_{0}) = \frac{1}{\mathcal{Z}}\textrm{tr}\{ I e^{-\beta \mathcal{H}}\}
\end{equation}
\noindent where $\mathcal{Z}$ is the partition function, $\mathcal{H}$ denotes the Hamiltonian and $\beta = 1/(k_{B}T)$ with $k_{B}$ being the Boltzmann constant. $I$ is the persistent current operator defined as
\begin{equation}\label{eq:current2}
    I = \frac{2\imath\pi t}{L}\sum\limits_{j=1}^{L}\sum\limits_{\alpha=1}^{N}\langle c_{j,\alpha}^{\dagger}c_{j+1,\alpha}e^{\frac{2\imath\pi}{L}\phi} - \textrm{h.c.} \rangle
\end{equation}
for a system of fermions with $N$ colours residing on a ring of $L$ sites with a hopping amplitude $t$ and pierced by an effective magnetic flux $\phi$.  \\
\begin{figure}[h!]
    \includegraphics[width=0.4\textwidth]{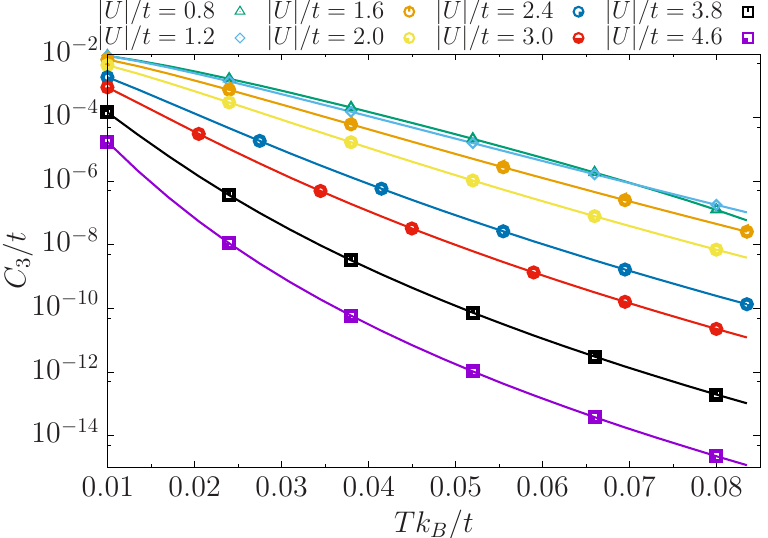}%
    \hspace{5mm}
    \includegraphics[width=0.4\textwidth]{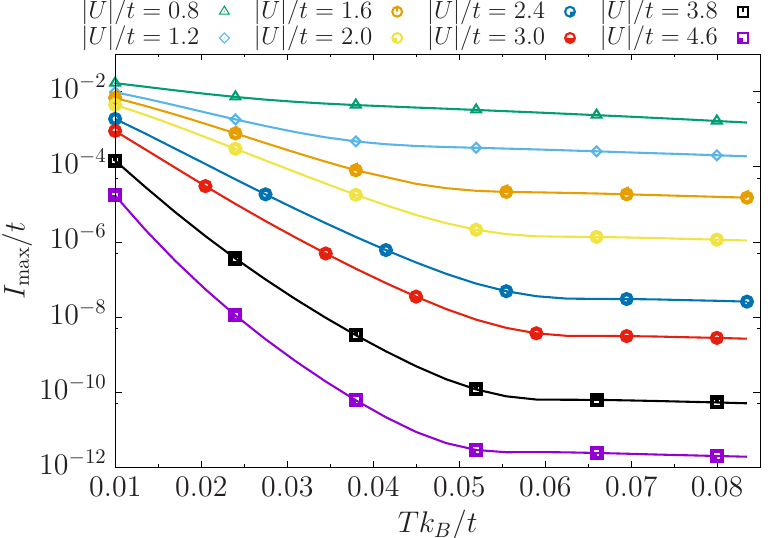}
    \put(-392,33){(\textbf{\textbf{a}})}
    \put(-170,33){(\textbf{\textbf{b}})}
    \caption{(\textbf{a}) Fourier weight $C_{3}/t$ as a function of temperature $Tk_{B}/t$ for different values of the interaction $U/t$. (\textbf{b}) depicts the corresponding maximum amplitude of the persistent current $I_{\mathrm{max}}/t$ against temperature $T k_{B}/t$ for different $U/t$. The presented results were obtained with exact diagonalization for $N_{p} = 3$ and $L=15$.  }
    \label{fig:fourier}
\end{figure}%

\begin{figure}[ht!]
    \includegraphics[width=\textwidth]{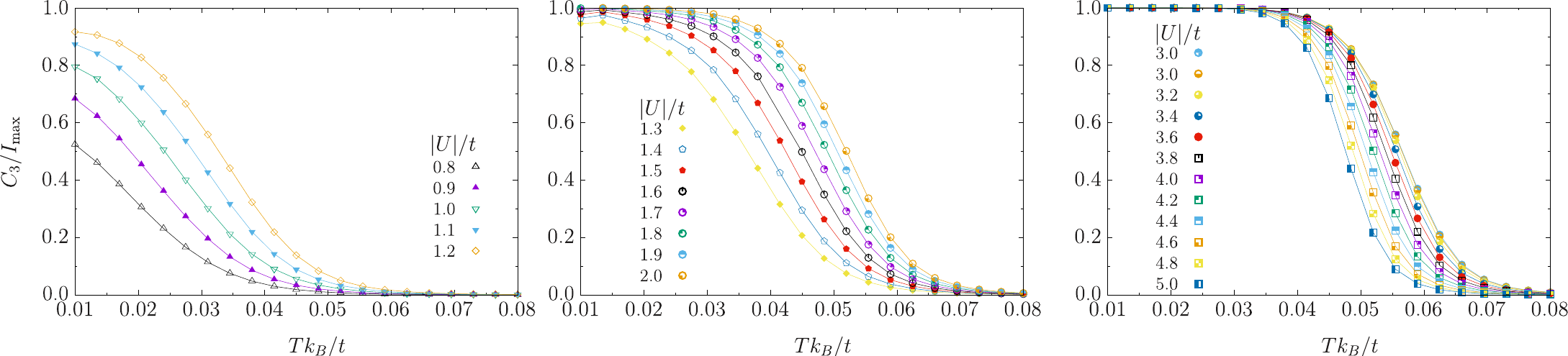}%
    \put(-360,95){(\textbf{\textbf{a}})}
    \put(-195,95){(\textbf{\textbf{b}})}
    \put(-21,95){(\textbf{\textbf{c}})}
    \caption{Figure of merit for the normalised Fourier weight $C_{3}/I_{\textrm{max}}$ as a function of temperature $T k_{B}/t$ for different values of the interaction $U/t$:
    (\textbf{a}) weak interactions; (\textbf{b}) intermediate interactions; and (\textbf{c}) strong interactions. The presented results were obtained with exact diagonalization for $N_{p} = 3$ and $L=15$. }
    \label{fig:raw}
\end{figure}

\noindent In our case, we follow the decay of the Fourier weight $C_{3}$ with temperature, since this weight corresponds to the formation of trions --Fig.~\ref{fig:fourier} (\textbf{a}). To normalise the Fourier weight, we use the maximum current amplitude $I_{\textrm{max}}$ at different temperature values --Fig.~\ref{fig:fourier} (\textbf{b}). Upon plotting the normalised Fourier weight, we notice three distinct regimes: regime (I) at weak values of the interaction where a trion has not yet reached a tri-partite periodicity Fig.~\ref{fig:raw} (\textbf{a}); regime (II) for the intermediate interaction where a trion is formed and there is an interplay between temperature and interaction Fig.~\ref{fig:raw} (\textbf{b}); and lastly regime (III) for strong interactions wherein we observe that increasing the interaction requires a lower value of the temperature to reinstate the persistent current with a single frequency Fig.~\ref{fig:raw} (\textbf{c}).\\

\begin{figure*}[h!]
    \includegraphics[width=0.45\textwidth]{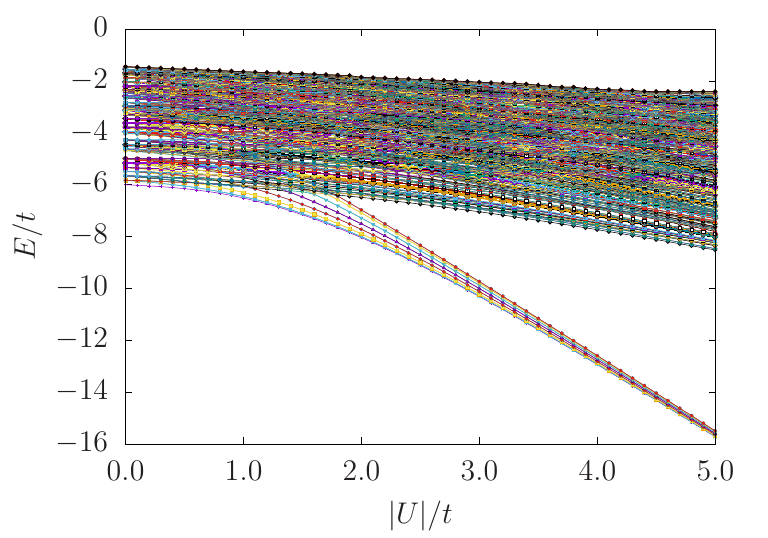}%
    \hspace{5mm}
    \includegraphics[width=0.45\textwidth]{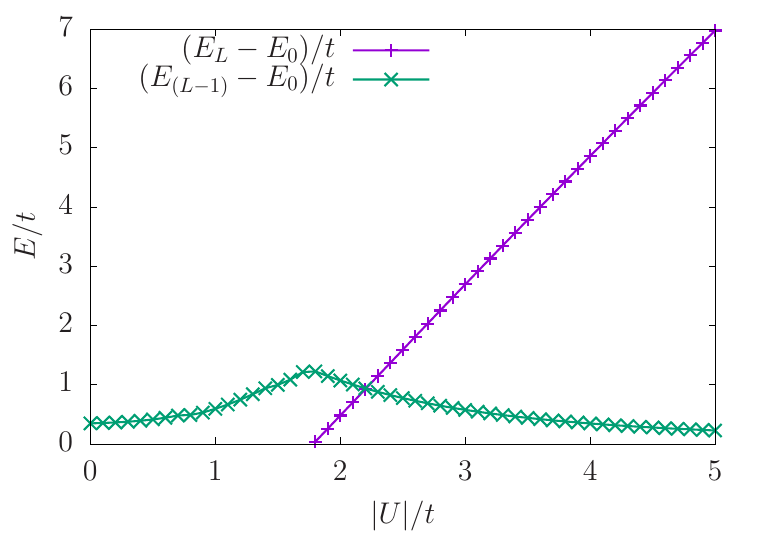}%
    \caption{(\textbf{a}) Energy spectrum $E/t$ as a function of the interaction $U/t$. For small and intermediate $U/t$, there is a continuous band between scattering and bound states. As $U/t$ is increased, a gap opens up that separates the bound and scattering states. Consequently for large $U/t$, the lower part of the spectrum is composed solely of bound states. (\textbf{b}) The green line depicts the separation in the lowest energy  band denoted as $(E_{L-1}-E_{0})/t$ where $E_{0}/t$ is the ground-state energy. The energy gap between the lowest band and the first excited state of the next band $(E_{L}-E_{L-1})/t$ is represented by the purple line. The presented results were obtained with exact diagonalization for $N_{p} = 3$ and $L=15$. }
    \label{fig:spec}
\end{figure*}

\newpage
\noindent The existence of these three regimes is corroborated by the energy spectrum --Fig.~\ref{fig:spec}. For small and moderate interactions, the system is characterized by a continuous band in which, beyond a certain energy threshold, the bound and scattering states are interwoven. For increasing attractions, scattering and bound states are organized in two distinct sub-bands separated by an increasing energy gap (linearly); the energy levels within the bound states sub-band result to be separated by a level spacing that is suppressed by interaction. \\

\noindent Below,  we describe the methodology that was employed to determine the functional forms  of the deconfinement phenomenon in the various regimes outlined  above. \\

\noindent The renormalised third Fourier coefficient of the current, denoted as $W = C_{3}/I_{\mathrm{max}}$, is a suitable indicator to track the change in the frequency of the persistent current as it goes from its trionic value three to a single particle frequency, which corresponds to when the first Fourier coefficient prevails. 
To identify the functional dependence of $W$ on the parameters, we employ the logic of the finite size scaling machinery~\cite{barber1983finite}. Our ansatz for $W$ is 
\begin{equation}\label{eq:scal1}
    (W-A_0)U^{\lambda} = G(T-T^{*})
\end{equation}
where  $T^{*}$ is a  crossover temperature defined by $W(T^*)=1/2$.
The value of  $\lambda$  is determined in such a way that a single functional law of the combination of  $U$ and $T$ is obtained. We observe that  $T^{*}$ depends non-monotonically on the interaction parameter $|U|$, indicating that the change of $W$ undergoes to distinct regimes.  \\

\noindent A parabolic behavior is observed in regimes I and II
\begin{equation}\label{eq:regime1&2}
T^*\approx T_{0,R} - a_{R}(U-U_{0,R})^2 
\end{equation}
where $T_{0,R}$ and  $U_{0,R}$ in the three regimes labelled by $R=\textrm{I,II,III}$ are fitting parameters.
In regime I, $T_{0,\textrm{I}}=0.040$, and $U_{0,\mathrm{I}}=1.61$, and  $a_{\mathrm{I}}=0.044$ are found. This value of $U_{0,\mathrm{I}}$ lies in the region where the spectrum starts to split (see Fig.~\ref{fig:spec}). Here, nearly all excited states are scattering states. \\

\begin{figure}[h!]
    \centering
    \includegraphics[width=0.4\linewidth]{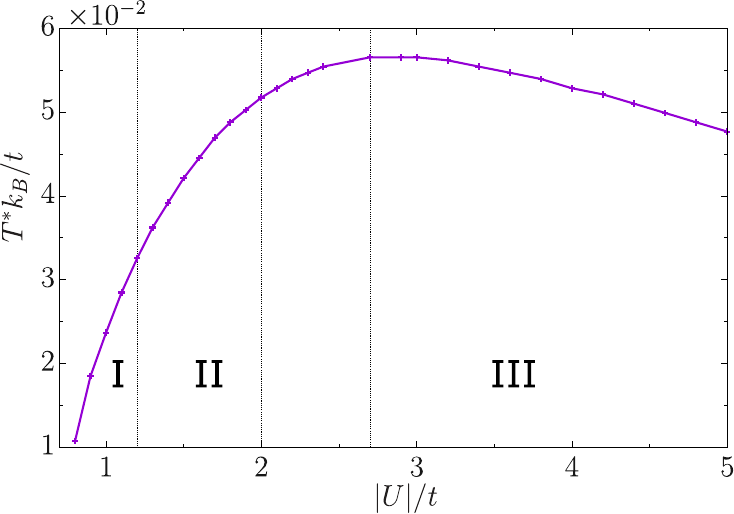}
    \caption{Crossover temperature $T^{*}k_{B}/t$ against interaction $U/t$ for  the three interaction regimes for: weak, intermediate and strong denoted as I, II and III respectively. Results were obtained with exact diagonalization.}
    \label{fig:tempc}
\end{figure}

\noindent In  regime II,  we also find this quadratic behavior
with  $T_{0,\mathrm{II}}=0.056$, $U_{0,\mathrm{II}}=2.56$, and $a_{II}=0.013$.
At this particular value of $|U|$, we experience that the maximum of $T^*$ is reached --Fig.~\ref{fig:tempc}. This regime corresponds to a clear distinction of the bound states from the scattering states. At the end of regime II, all avoided level crossings between scattering states and bound states are resolved, with the two bands becoming fully separate.  \\

\noindent In regime III, the behavior is instead fitted by
\begin{equation}\label{eq:regime 3}
T^*\approx T_{0,\mathrm{III}}-a_{\mathrm{III}}(U-U_{0,\mathrm{III}})^{1.459}
\end{equation}
choosing the same value of $|U_{0,\mathrm{III}}|=|U_{0,\mathrm{II}}|$.
We find   $T_{0,\mathrm{III}}=0.0576=T_{0,\mathrm{II}}$, together with $a_{\mathrm{III}}=0.0027$. This regime corresponds to a clear distinction of the bound states from the scattering states. Note that despite the small temperatures considered here, all bound states belonging to the lowest band are involved in the dynamics. \\

\noindent Inserting the respective values for all regimes into a Taylor expansion for the function $G$ demonstrates 
that in the vicinity of the crossover $|U-U_{0,\mathrm{II}}|<2.5$,
no impact is made on final form of the function G, especially near $T^*$.
Upon re-arrangement we have that
\begin{equation}\label{eq:scal2}
    W-A_0 = g_{R}(U) G_R\bigg(\frac{T-T_{0,R}}{(U-U_{0,R})^{\mu_{R}}}\bigg)
\end{equation}
where $g_{\mathrm{I}}=U^{1.25}(U-U_{0,\mathrm{I}})^2$, $\mu_{\mathrm{I}}=2$, $g_{\mathrm{II}}=U^{0.33}(U-U_{0,\mathrm{II}})^2$, $\mu_{\mathrm{II}}=2$, and $g_{\mathrm{III}}=U^{0.1}(U-U_{0,\mathrm{III}})^{1.459}$, $\mu_{\mathrm{III}}=1.459$. 
In this last region, all the bound states are nearly degenerate and any small temperature is relevant to combine all the contributions of the current in the excited states (we note that the  scattering states are well separated here from the bound states sub-band --see Fig.~\ref{fig:spec}). \\

\noindent Lastly, we would like to comment on the finite temperature effects on the persistent current of a system in a CSF configuration. The persistent currrent frequency indicates that, compared with trions, CSFs are less robust  to thermal fluctuations. However, looking at the persistent current per colour paints a more interesting picture. The total current has a period given by the bare flux quantum purely because the persistent current of the two colours in the pair is smaller than that of the unpaired colour. Interestingly enough, the temperature required to break the interaction between the pair is higher than that required for a system of symmetric trions with the same interaction. A proper study on the finite temperature effects on the persistent current of a CSF will be done in a separate work.

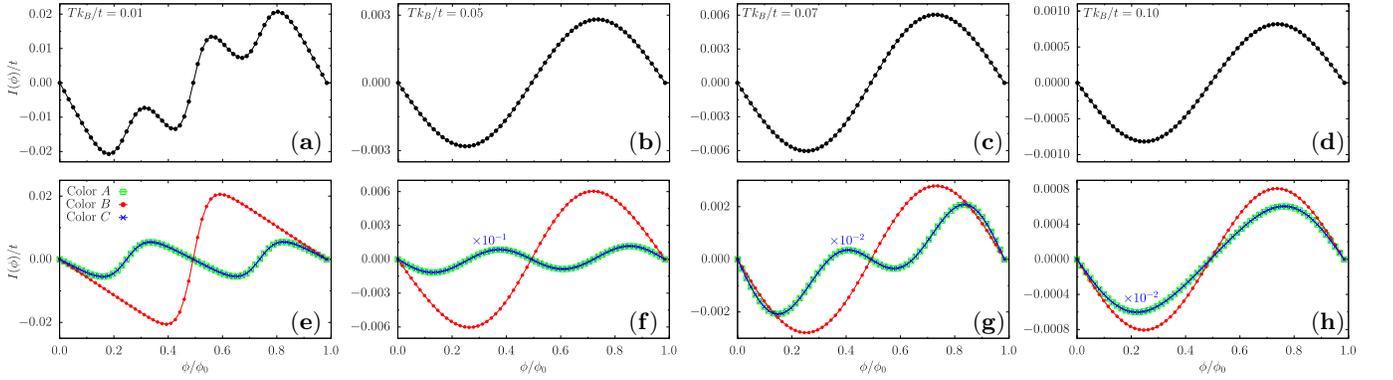
\begin{figure*}[h!]
    \includegraphics[width=1\textwidth]{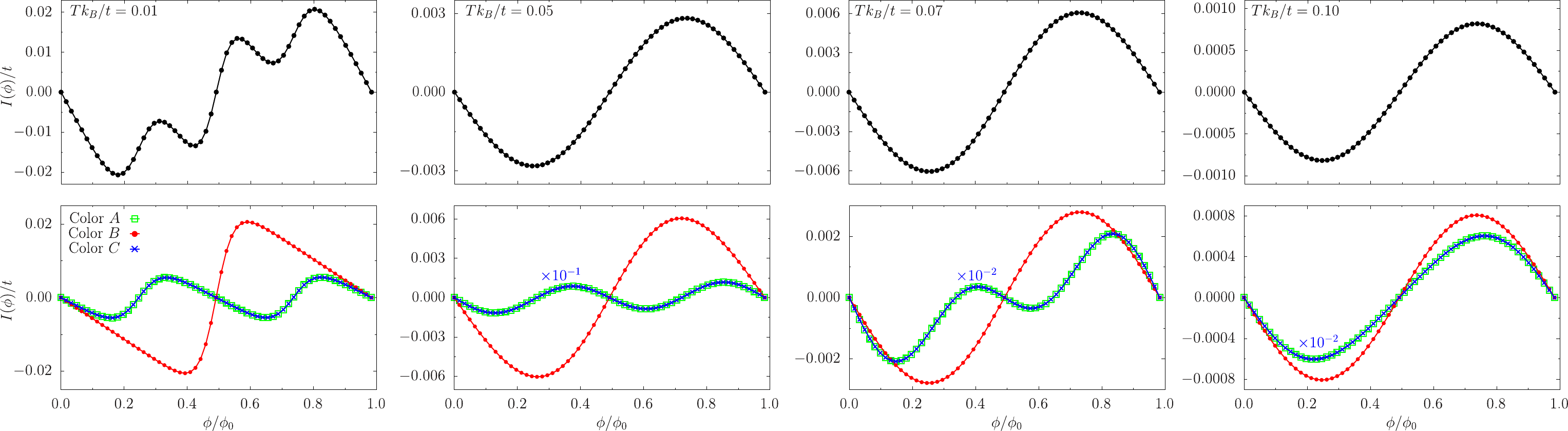}%
    \put(-403,85){(\textbf{\textbf{a}})}
    \put(-275,85){(\textbf{\textbf{b}})}
    \put(-145,85){(\textbf{\textbf{c}})}
    \put(-17,85){(\textbf{\textbf{d}})}
    \put(-403,17){(\textbf{\textbf{e}})}
    \put(-275,17){(\textbf{\textbf{f}})}
    \put(-145,17){(\textbf{\textbf{g}})}
    \put(-17,17){(\textbf{\textbf{h}})}
    \caption{Top (bottom) panels depict the persistent current $I(\phi)/t$ for all (each) colours against the flux $\phi/\phi_{0}$ for different values of the temperature $T k_{B}/t$. All the results were obtained with exact diagonalization for a system of $N_{p}=3$ and  $L=20$ in a CSF configuration for $|U_{AB}|/t = |U_{BC}|/t=0.01$ and $|U_{AC}|/t=3$. The lines  are meant as a guide to the eye for the reader.}
    \label{fig:csftemp}
\end{figure*}

\end{appendix}

\end{document}